\documentclass[preprint2]{aastex}

\newcommand{\lea}{{\>\rlap{\raise2pt\hbox{$<$}}\lower3pt\hbox{$\sim$} \>}}
\newcommand{\gea}{{\>\rlap{\raise2pt\hbox{$>$}}\lower3pt\hbox{$\sim$} \>}}
\begin{document}

%\title{NEW TESTS FOR MASS-DEPENDENT DISRUPTION OF STAR CLUSTERS:
%        FIRST APPLICATION TO THE ANTENNAE GALAXIES}

\title{NEW TESTS FOR DISRUPTION MECHANISMS OF STAR CLUSTERS: METHODS AND APPLICATION TO THE ANTENNAE GALAXIES}

\author{S. Michael Fall     \altaffilmark{1,2},
         Rupali Chandar      \altaffilmark{3}, and
         Bradley C. Whitmore \altaffilmark{2}
         }

%% are identified by the \altaffilmark after each name.  Specify  
%%alternate

\altaffiltext{1}{Institute for Advanced Study,
         Einstein Drive, Princeton, NJ 08540}
\altaffiltext{2}{Space Telescope Science Institute,
         3700 San Martin Drive, Baltimore, MD 21218;
         fall@stsci.edu, whitmore@stsci.edu}
\altaffiltext{3}{Department of Physics and Astronomy,
         University of Toledo, Toledo, OH 43606;
         rupali.chandar@utoledo.edu}

\begin{abstract}

We present new tests for disruption mechanisms of star 
clusters based on the bivariate mass-age distribution
$g(M, \tau)$. 
In particular, we derive formulae for $g(M,\tau)$ for
two idealized models in which the rate of disruption
depends on the masses of the clusters and one in which
it does not.  
We then compare these models with our {\it Hubble 
Space Telescope\/} observations of star clusters in the 
Antennae galaxies over the mass-age domain in which we 
can readily distinguish clusters from individual stars: 
$\tau \la 10^7 (M/10^4 M_{\odot})^{1.3}$~yr.
We find that the models with mass-dependent disruption
are poor fits to the data, even with complete freedom 
to adjust several parameters, while the model with 
mass-{\it in\/}dependent disruption is a good
fit.
The successful model has the simple form $g(M,\tau) 
\propto M^{-2} \tau^{-1}$, with power-law mass and
age distributions, $dN/dM \propto M^{-2}$ and 
$dN/d\tau \propto \tau^{-1}$.
The predicted luminosity function is also a power law, 
$dN/dL \propto L^{-2}$, in good agreement with 
our observations of the Antennae clusters.
The similarity of the mass functions of star clusters
and molecular clouds indicates that the efficiency of 
star formation in the clouds is roughly independent of 
their masses.
The age distribution of the massive young clusters is
plausibly explained by the following combination of
disruption mechanisms: (1) removal of interstellar
material by stellar
feedback, $\tau \la 10^7$~yr; (2) continued stellar
mass loss, $10^7 {\rm yr} \la \tau \la 10^8 {\rm yr}$;
(3), tidal disturbances by passing molecular clouds,
$\tau \ga 10^8$~yr.
None of these processes is expected to have a strong
dependence on mass, consistent with our observations
of the Antennae clusters.
We speculate that this simple picture also 
applies---at least approximately---to the clusters 
in many other galaxies.

\end{abstract}


\section{INTRODUCTION}

Star clusters are born in molecular clouds and are then
dispersed into the surrounding stellar field by a variety
of processes operating on different timescales, including
the removal of interstellar material by stellar feedback,
continued stellar mass loss, tidal disturbances by passing
molecular clouds, gravitational shocks during rapid passages
near the galactic bulge or through the galactic disk,
orbital decay into the galactic center caused by dynamical
friction, and stellar escape driven by internal two-body
relaxation.
Signatures of these processes are encoded in the mass
function, $\psi(M) \propto dN/dM$, and the age distribution,
$\chi(\tau) \propto dN/d\tau$, for a population of clusters
in a particular galaxy.
A more informative, but less familiar statistic is the
bivariate distribution of masses and ages, $g(M,\tau) \propto
\partial^2N/\partial{M}\partial{\tau}$.
The key feature of $g(M,\tau)$ is that it includes correlations
between $M$ and $\tau$, information that is absent from the
univariate distributions $\psi(M)$ and $\chi(\tau)$.
Thus, with the help of $g(M, \tau)$, we can answer questions
such as whether low-mass clusters are disrupted faster than,
or at the same rate as, high-mass clusters.
In this paper, we derive formulae for $g(M, \tau)$ for the
first time for several different models for the disruption
of clusters, and we then compare these formulae with our
{\it Hubble Space Telescope\/} ({\it HST\/}) observations of clusters 
in the Antennae galaxies.
In a companion paper, we present similar comparisons for the
clusters in the Large and Small Magellanic Clouds (LMC and
SMC; Chandar et al. 2009, hereafter CFW09).

We have focused on the star clusters in the Antennae galaxies
for several reasons.
First, the population of clusters is large, $N \approx 2300$
brighter than $M_V = -9$, ample for statistical studies.
Second, many of the bright young clusters have masses in the
range of old globular clusters, $10^4 \la M/M_{\odot} \la 10^6$,
suggesting that the former may simply be younger versions of
the latter.
Third, because the Antennae are currently in the throes of a
major merger, they give us a close-up, internal view of events
and processes that were more common in galaxies in the past,
during their hierarchical formation.
Fourth, the Antennae have been mapped at almost every wavelength
available to astronomers, from radio to X-rays, providing a more
detailed picture of their stellar and interstellar contents than
for almost any other galaxy outside the Local Group.

In our previous studies of the clusters in the Antennae, we
found that the mass and age distributions could be approximated
by power laws: $\psi(M) \propto M^{\beta}$ with $\beta \approx -2$
(Zhang \& Fall 1999, hereafter ZF99) and $\chi(\tau) \propto
\tau^{\gamma}$ with $\gamma \approx -1$ (Fall et al. 2005,
hereafter FCW05).
In these studies, we also found that the shapes of $\psi(M)$
and $\chi(\tau)$ are approximately independent of the age and
mass limits, respectively, of the samples used to determine them.
These findings suggest that the bivariate mass-age distribution
can be approximated simply by the product of the univariate
distributions:
\begin{equation}
g(M, \tau) \propto \psi(M)\chi(\tau) \propto M^{\beta}\tau^{\gamma}
\end{equation}
\begin{equation}
{\rm for} \,\,\,\,\,\,\,\,\,
\tau \la 10^7 (M/10^4 M_{\odot})^{1.3}\,{\rm yr}.
\end{equation}
The second expression above indicates the approximate domain
of validity of the first, the condition that objects in the
sample be brighter than the most luminous individual stars
($L_V \ga 3\times10^5 L_{\odot}$).\footnote
{Equation~(2) follows from the fact that the mass-to-light
ratios of clusters vary with age approximately as
$M/L_V \propto \tau^{0.8}$ for $\tau \ga 10^7$~yr.}
In this model, clusters form with a power-law initial mass
function and are then disrupted at rates that are independent
of their masses.
We have already explored some of the consequences of equation~(1)
in two recent papers (Fall 2006; Whitmore et al. 2007, hereafter
WCF07).
In this paper, we make further tests of this model, and
we also compare our {\it HST\/} observations of the Antennae 
clusters with two models in which clusters of different 
masses are disrupted at different rates.

The age distribution for a population of star clusters reflects,
in principle, a combination of both formation and disruption rates.
We interpret the decline of $\chi(\tau)$ primarily as the result 
of disruption rather than formation for the following reasons.
First, the age distribution has a sharp peak
at the present time, $\tau = 0$, and a small width,
characterized by the median age $\tau \sim 10^7$~yr.
This requires either rapid disruption at an arbitrary time
(now) or rapid formation at a special time (now), the former
being much more likely a priori than the latter.
Of course, the Antennae galaxies are presently merging, which
has almost certainly boosted the formation rates of stars and
clusters, but this occurs more slowly, on the orbital timescale 
of the galaxies, $\tau \sim {\rm few} \times 10^8$~yr. 
In the simulations of merging galaxies by Mihos et al. (1993), 
including the Antennae, for example, the star formation rate
varies by factors of only a few or less in the past $10^8$~yr,
whereas the observed age distribution of the clusters drops by
nearly 2 orders of magnitude in this interval of time. 
The second reason we interpret the decline of $\chi(\tau)$ 
in terms of disruption is that it has the same shape, 
including the sharp peak at $\tau = 0$,  
in large regions separated by distances of order 10~kpc 
within the Antennae galaxies (Whitmore 2004; WCF07).
There is no physical mechanism that could synchronize a
burst of cluster formation this precisely at these separations;
the communication time is $\tau \sim 10^9$~yr for a signal
traveling at the typical random velocity in the interstellar
medium (ISM; $v \sim 10$~km s$^{-1}$) and $\tau \sim 10^8$~yr 
for one traveling at the highest bulk velocity ($v \sim
100$~km s$^{-1}$).
Thus, we conclude that the observed decline in the age
distribution is caused mainly by disruption, at least for
$\tau \la \mbox{few}\times10^8$~yr.

We have speculated that equation~(1) for $g(M, \tau)$ 
may also describe---approximately---the cluster 
populations in other galaxies (Fall 2006; WCF07).
If this picture turns out to be generally valid, it will
mean that star clusters form and evolve in much the same
way in different galaxies, the primary variable being an
overall scale factor proportional to the total number of
clusters and hence to their average formation rate.
We define a cluster here to be any concentrated
aggregate of stars with a density much higher than that
of the surrounding stellar field, whether or not it is
gravitationally bound, since this is virtually impossible
to determine from the available observations, especially
for clusters younger than about 10 internal crossing times.
Most, if not all, stars form in clusters (as just defined)
and most clusters then dissolve into the stellar field.
We find it intriguing that this whole complex process might
be represented approximately by a simple ``recipe'' such
as equation~(1).

There are a few other hints that point toward this model:
(1) The same mass and age distributions, $\psi(M) \propto
M^{-2}$ and $\chi(\tau) \propto \tau^{-1}$, are found for
clusters with $M \la 10^3 M_{\odot}$ and $\tau \la
3 \times 10^8$~yr in the solar neighborhood (Lada \& Lada
2003).
(2) The luminosities of the brightest clusters in different
galaxies scale approximately with the size of the population
in the way expected from equation~(1) (Larsen 2002; Whitmore
2003; WCF07).
(3) The mass spectrum of molecular clouds, from which the
clusters form, appears to be similar in different galaxies
(Blitz et al. 2007).
(4) The most important early disruption processes, removal
of ISM by stellar feedback and continued stellar mass loss,
are independent of the properties of the host galaxy (except
possibly its metallicity and stellar initial mass function 
(IMF)).
Of course, a definitive conclusion about the generality
of equation~(1) will require more detailed studies, such
as that presented here for the Antennae, of the cluster
populations in several other galaxies.
We have recently made a start on this with the LMC and
SMC (CFW09).
Our main purpose in this paper is to present some of the
interpretive tools needed for such studies.

The plan for the remainder of this paper is the following:
In \S2, we present analytic formulae for $g(M, \tau)$ for
two models with mass-dependent disruption and the
model already mentioned with mass-independent disruption.
In \S3, we describe our {\it HST\/} observations of the Antennae
clusters and compare these with the models presented in the
previous section.
We also derive the luminosity function of the Antennae
clusters and show how this is related to the mass function
through $g(M, \tau)$.
In \S4, we discuss the physical processes most likely
responsible for the formation and disruption of clusters
and how these processes affect the mass and age distributions.
This interpretative section presents our current understanding
of the entire life cycles of star clusters.
In \S5, we summarize our main results and their broader
implications.
The paper also includes two appendices.
The first (A) discusses some related work on the mass and                     
luminosity functions of the Antennae clusters by Mengel
et al. (2005) and by Anders et al. (2007) respectively,
while the second (B) presents some specialized
formulae needed in \S3.

\section{MODELS}

The basic element of a statistical description of
a population of star clusters is the multivariate
distribution of all the important properties of the
clusters, such as their luminosities, masses, ages,
effective radii, internal concentrations, stellar
contents, and so forth.
For a population of clusters more distant than $\sim10$
Mpc, however, many clusters are poorly resolved
or not resolved at all, and the list of available properties
shrinks to just three: luminosities, masses, and ages, of
which only two are independent if the stellar IMF is 
assumed to be universal, as it almost always
is in practice.
With these thoughts in mind, we focus here on the bivariate
distribution $g(M, \tau)$, defined such that $g(M, \tau)
dMd\tau$ is the number of clusters with masses between
$M$ and $M+dM$ and ages between $\tau$ and $\tau+ d\tau$.
This distribution can be estimated from fits of stellar 
population models to photometry in several bands of the 
clusters in a large sample, as we describe in the next 
section.

The goal of this section is to relate $g(M, \tau)$ to
information about the formation and disruption of the
clusters.
This is done most easily with the help of a few auxiliary
functions.
We define another distribution $f(M_0, \tau)$ such that
$f(M_0, \tau)dM_0d\tau$ is the number of clusters born
with initial masses between $M_0$ and $M_0+dM_0$ at times
in the past between $\tau$ and $\tau+d\tau$.
Furthermore, we assume that the current mass $M(M_0, \tau)$
of a cluster with an initial mass $M_0$ and the inverse
relation $M_0(M, \tau)$ are known functions of the age $\tau$.
Then $f$ and $g$ are related by the continuity equation
in the form
\begin{equation}
g(M,\tau) = f[M_0(M, \tau), \tau]
(\partial{M_0}/\partial{M})_{\tau}.
\end{equation}
This generalizes equation~(2) of Fall \& Zhang (2001) from
an instantaneous burst to an arbitrary history of cluster
formation.
In equation~(3) here, all information about formation
is encoded in $f(M_0, \tau)$, while all information about
disruption is encoded in $M(M_0, \tau)$ and hence in the
difference between $f$ and $g$.
In fact, these distributions are equal only if clusters
are never disrupted (i.e., $M = M_0$ for all $\tau$).

We can exploit equation~(3) in either of two ways.
If we knew everything about the disruption mechanisms
(likely more than one) and hence the precise form of
$M(M_0, \tau)$, we could use the empirical determination
of $g(M, \tau)$ together with equation~(3) to infer
$f(M_0, \tau)$.
In practice, however, we must work in the other direction:
make reasonable assumptions about $f(M_0, \tau)$, adopt a
specific model for $M(M_0, \tau)$, compute $g(M, \tau)$
from equation~(3), and compare this with the corresponding
empirical distribution to see if the model is acceptable
or must be rejected.
In this spirit, we assume that clusters form with the
following initial mass-age distribution:
\begin{equation}
f(M_0, \tau) = C(\tau) (M_0/M_*)^{\beta}.
\end{equation}
Here $C(\tau)$, with units of [mass]$^{-1}$~[time]$^{-1}$, 
specifies the formation rate of clusters at an age $\tau$,
while $M_*$ is a fiducial mass, usually taken to be 
$10^4~M_{\odot}$.
Our assumption that the initial mass function is a 
power law is justified by our previous results and 
those presented in the next section of this paper.
In all our detailed comparisons with observations,
we assume for simplicity that the formation rate 
$C(\tau)$ in equation~(4) is constant.
For the reasons discussed in the Introduction, this  
should be a good approximation for the Antennae 
clusters younger than a $\mbox{few}\times10^8$~yr.
It is worth noting here, however, that the mathematical
formulae for $g(M, \tau)$ presented in this section 
[equations~(6), (12), and (13) below] are equally valid 
with a variable formation rate, even one as extreme as 
an instantaneous burst, with 
$C(\tau) \propto \delta (\tau - \tau_0)$.
We also note that, although $f(M_0, \tau)$ is a separable
function of $M_0$ and $\tau$ (by assumption), $g(M, \tau)$
will generally include correlations between $M$ and $\tau$,
except in the special case in which $M/M_0$ is a function
only of $\tau$.

In the remainder of this section, we present the mass-age
distribution $g(M, \tau)$ for three simple, illustrative
models for the disruption of clusters. 
In this context, it is important to bear in mind that
several processes probably operate together to disrupt
the clusters, with different combinations predominating
at different ages over the lifetimes of the clusters.
Thus, we cannot expect the most successful model to 
display the signature of any single process operating 
in isolation.
The massive young clusters in our
sample are probably disrupted first by the
removal of residual interstellar material driven by
early stellar feedback (photoionization, 
radiation pressure, stellar winds, and
supernovae), then by continued mass loss from
the stars themselves through stellar winds and other
ejecta, and then
possibly by the escape of stars driven by tidal
disturbances of passing structures such as molecular
clouds and/or spiral arms.
The three models we consider have simple functional forms 
but several adjustable parameters to accommodate this likely 
complexity.
We note here that the clusters in our 
sample are all too massive and too young to have lost a 
significant fraction of their mass as the result of stellar
escape driven by internal two-body relaxation, the primary 
long-term disruptive process for clusters 
and one that certainly depends 
on mass (see \S 4.3 below).

{\it Model~1: Sudden Mass-Dependent Disruption\/}.
In this case, we assume clusters retain all of
their initial mass until they are destroyed
suddenly and completely at an age $\tau_d(M_0)$:
\begin{equation}
M(\tau) = \left\{
   \begin{array}{ll}
     M_0 & \mbox{for $\tau \le \tau_d(M_0)$}
\\
     0 & \mbox{otherwise.}
   \end{array} \right.
\end{equation}
The corresponding mass-age distribution, from
equations~(3), (4), and (5), is
\begin{equation}
g(M, \tau) = \left\{
   \begin{array}{ll}
     C(M/M_*)^{\beta} & \mbox{for $\tau \le \tau_d(M)$}
\\
     0 & \mbox{otherwise.}
   \end{array} \right.
\end{equation}
These equations are valid for any dependence of
$\tau_d(M_0)$ on $M_0$.
Boutloukos \& Lamers (2003) have proposed a specific
version of this model with a power-law dependence on mass:
\begin{equation}
\tau_d(M_0) = \tau_* (M_0/M_*)^k.
\end{equation}
The discontinuous evolution posited by this model
is clearly an oversimplification.
Boutloukos \& Lamers (2003) claim that Model~1 does, 
nevertheless, reveal the correct mass dependence of 
disruption when it is compared with observations.
They find the same exponent $k \approx 0.6$ but very 
different disruption timescales $\tau_*$ for the 
cluster populations in several galaxies.
We are skeptical of these claims for reasons we explain
in detail in \S3 and \S4 (see also WCF07).

{\it Model~2: Gradual Mass-Dependent Disruption\/}.
In this case, we assume clusters lose mass gradually
according to the equations
\begin{equation}
dM/d\tau = - M/\tau_d(M),
\end{equation}
\begin{equation}
\tau_d(M) = \tau_* (M/M_*)^k.
\end{equation}
The similarity between the disruption timescales
$\tau_d(M_0)$ and $\tau_d(M)$ in equations~(7) and~(9)
makes Model~2 an analog of Model~1 but with continuous
evolution (as noted also by Boutloukos \& Lamers 2003).
It seems reasonable to suppose that the gradual 
evolution posited by Model 2 would make it more realistic 
and hence a better match to the observations than Model~1.
Indeed, Model~2 has known physical justifications 
in two special cases, both of which apply to tidally 
limited clusters (with constant mean density).
For $k = 0$, it describes the disruption of clusters
by external gravitational shocks, while for $k = 1$,
it describes the disruption of clusters by internal
two-body relaxation (Spitzer 1987).
The first of these is potentially relevant to 
intermediate-age clusters, while the second is 
certainly relevant to old clusters, as we discuss
in detail in \S4. 

The solutions of equations~(8) and~(9) are
\begin{equation}
M(\tau) = \left\{
   \begin{array}{ll}
     M_0 \left[1 - k (M_0/M_*)^{-k}(\tau/\tau_*)\right]^{1/k}\\
     \qquad \mbox{for $k \not= 0$}
\\
     M_0 \exp (-\tau/\tau_*)\\
     \qquad \mbox{for $k = 0$.}
   \end{array} \right.
\end{equation}
Thus, the derivative required in equation~(3) is
\begin{equation}
(\partial{M_0}/\partial{M})_{\tau} = (M_0/M)^{1-k}
\hspace{1cm} {\rm for} \,\, {\rm all} \,\, k.
\end{equation}
Combining equations~(3), (4), (10), and (11), we obtain
the mass-age distribution
\begin{equation}
g(M, \tau) = \left\{
   \begin{array}{ll}
     C (M/M_*)^{\beta}\left[1 + k(M/M_*)^{-k}
     (\tau/\tau_*)\right]^{(\beta + 1 - k)/k}\\
      \qquad\mbox{for $k \not= 0$}
\\
     C (M/M_*)^{\beta}
     \exp\left[(\beta + 1)(\tau/\tau_*)\right]\\
     \qquad \mbox{for $k = 0$.}
   \end{array} \right.
\end{equation}
Several features of this equation are noteworthy.
For fixed $\tau$ and $k > 0$, $g(M, \tau)$ is
a double power law with a low-$M$ exponent $k - 1$,
a gradual bend at the mass given by $\tau \approx \tau_d(M)$,
and a high-$M$ exponent $\beta$.
For fixed $M$ and $k \not= 0$, $g(M, \tau)$ is constant
for small $\tau$, has a knee at $\tau \approx \tau_d(M)$,
and is a power law with an exponent $(\beta + 1 - k)/k$
for large $\tau$.
However, for $k = 0$, $g(M, \tau)$ is a single power
law in $M$ with an exponent $\beta$ for fixed $\tau$,
and declines exponentially in $\tau$ with a timescale
$-\tau_*/(\beta + 1)$ (i.e., faster than any power law)
for fixed $M$.

{\it Model~3: Gradual Mass-Independent Disruption\/}.
In this case, we adopt the mass-age distribution
indicated by our earlier studies (ZF99, FCW05, Fall 2006, WCF07),
namely
\begin{equation}
g(M,\tau) = C(M/M_*)^{\beta}(\tau/\tau_*)^{\gamma}.
\end{equation}
This distribution is manifestly separable: 
$g(M, \tau) \propto \psi(M)\chi(\tau)$, with $\psi(M) \propto 
(M/M_*)^{\beta}$ (independent of $\tau$) and $\chi(\tau) 
\propto (\tau/\tau_*)^{\gamma}$ (independent of $M$).
It has three independent, adjustable parameters: $C$,
$\beta$, and $\gamma$.
The other parameters, $M_*$ and $\tau_*$ (which are not 
independent of $C$, $\beta$, and $\gamma$), merely 
specify the units in which masses and ages are 
measured.\footnote{Note that the parameter $\tau_*$ plays
different roles in the three models; for Models~1 and 2,
it specifies the characteristic disruption timescale
for clusters of mass $M_*$, while for Model~3, it 
enters only as an arbitrary normalization factor.}
We can derive equation~(13) from equations~(3) and~(4)
with $M(\tau) = M_0(\tau/\tau_*)^{-\gamma/(\beta + 1)}$.
In this case, the masses of all clusters decline gradually,
with the same power law in age.
However, this interpretation of equation~(13) is not unique.
Another possibility is that $g(M, \tau)$ specifies the 
probability that clusters of mass $M$ survive to an age 
$\tau$, even if disruption, when it happens, is sudden.
The form of $g(M, \tau)$ displayed in equation~(13) then
implies that the survival probability is independent
of mass and declines as a power law in age.
Intermediate cases are also possible, with different 
combinations of age-dependent masses and survival
probabilities.
Moreover, some clusters may lose only part of their mass
(generally in the form of the least bound stars), while 
others are completely destroyed.
We refer to all situations described by equation~(13) as 
gradual mass-independent disruption, irrespective of
whether the disruption of individual clusters is gradual 
or sudden, partial or complete, because it leads to a 
gradual decline in the number of clusters of each mass
with age at a fractional rate independent of mass.\footnote
{When we mention rates in this paper, we usually mean
fractional rates, i.e., $d(\ln X)/dt$ rather than $dX/dt$.}

The main justification for Model~3 is that it provides
a simple, approximate representation of $g(M,\tau)$ for 
the massive young clusters in the Antennae galaxies (in 
the mass-age domain specified by eqn.~[2] above).  
It is likely that the power-law dependence of $g(M,\tau)$ on
$M$ reflects the mass function of the molecular clouds in which the
clusters formed, and an efficiency of star formation that is nearly
independent of the masses of the clouds.
The subsequent disruption of many of the young clusters
by the removal of their ISM, stellar mass loss, and tidal disturbances
also tends to preserve the power-law shape of the mass function.
These effects plausibly predominate in the following sequence:
(1) removal of ISM, $\tau \la 10^7$~yr;
(2) stellar mass loss, $10^7 {\rm yr} \la \tau \la 10^8 {\rm yr}$; 
(3) tidal disturbances, $\tau \ga 10^8$~yr.  
This combination of disruptive processes might then account
(approximately) for the power-law dependence of $g(M,\tau)$
on $\tau$.
Variations in the formation rate could also play some role,
although we expect this to be minor, for the reasons discussed in 
the Introduction.
We discuss these issues more fully in \S4.

\section{OBSERVATIONS AND COMPARISONS WITH MODELS}

The goal of this section is to compare the observed
mass and age distributions of star clusters in the Antennae
with predictions from the three disruption models developed
in the previous section.
In the next section, we discuss the physical implications of 
our results for the formation and disruption of the clusters, 
and in a companion paper we make similar comparisons between 
models and data for clusters in the Magellanic Clouds (CFW09).

\subsection{Observations and Input Data}

We have derived the masses and ages of the Antennae clusters 
from \textit{UBVI}H$\alpha$ photometry of images taken with the 
WFPC2 on {\it HST\/}.
These images cover nearly all of the main bodies of the
Antennae galaxies, omitting only a small region in the northwest
and the long tidal tails.
Thus, all the results presented in this paper are essentially 
averages over the two merging galaxies.
We describe our procedure in detail in FCW05, including extensive
tests of its validity, and the resulting age distributions for
mass- and luminosity-limited samples of clusters.  
Here, we give only a brief summary of our procedure but a more
complete set of results, including the bivariate mass-age 
distribution and the univariate mass and age distributions.
The present work also updates the earlier results from ZF99 based 
on \textit{UBVI} photometry of the same images, but with masses and
ages estimated from two reddening-free $Q$ parameters.

In the present study, we determine the age $\tau$ and extinction 
$A_V$ of each object by performing a minimum $\chi^2$ fit between
the observed \textit{UBVI}H$\alpha$ magnitudes and those predicted
by stellar population models. 
We use the Bruzual \& Charlot (2003) models with 
solar metallicity, a Salpeter (1955) IMF
with lower and upper cutoffs at 0.1~$M_{\odot}$ and 100~$M_{\odot}$,
and the Galactic-type extinction law from Fitzpatrick (1999).
Our procedure, which uses measurements in five filters (four 
colors) simultaneously to estimate two parameters ($\tau$ and $A_V$)
for each cluster, is thus over-constrained mathematically.
The H$\alpha$ flux varies rapidly with age near $\approx10^7$~yr,
and is particularly useful in distinguishing clusters younger 
and older than this. 
We estimate the masses of the clusters from their total
$V$-band luminosity (corrected for extinction) and the age-dependent
mass-to-light ratio ($M/L_V$) from the stellar population models, 
assuming a distance of 19.2~Mpc to the Antennae.
Figure~1 shows the resulting luminosity-age distribution for all 
objects, and Figure~2 shows the equivalent mass-age distribution.
If, instead of the Salpeter (1955) IMF, we had adopted the Kroupa
(2001) or Chabrier (2003) IMF, all the masses would be reduced by
about 40\% while all the ages would remain virtually the same 
(see Fig.~1 of CFW09).

The objects plotted in Figures~1 and 2 include both individual
stars and compact star clusters.
If the Antennae were closer we could use
structural parameters alone to select the clusters,
but not all of them
are spatially resolved in the $HST$ images (i.e. broader
than the point-spread function), 
and differentiating clusters from individual stars becomes
increasingly difficult at faint magnitudes and in 
crowded regions.
Therefore,
following the standard technique (e.g., Whitmore et al. 1999),
we minimize contamination by restricting
our sample to objects brighter than all but the most
luminous stars ($L_V \geq 3 \times 10^5 L_{\odot}$
or $M_V\leq-9$), resulting in stellar
contamination at the
$\la5$\% level (ZF99), and
giving a final sample of $\approx2300$ clusters.
The solid lines in Figures~1 and 2 show
our cluster selection limit of $M_V=-9$.
While our optically selected catalog undoubtedly
misses some very young, dust-enshrouded clusters,
we estimate it to be complete at the $\ga85$\% level for
clusters with $\tau\la10^7$~yr, based on positional coincidences
between radio-continuum and optically selected sources
(Whitmore \& Zhang 2002), and between infrared and optically
selected sources (Whitmore et al. 2009).

We have performed a variety of tests to assess the reliability
of our age estimates.
These include repeating the entire analysis
using both the Bruzual \& Charlot (2003) and the Starburst99 (Leitherer et al. 1999) 
stellar population models, including and excluding
the H$\alpha$ measurements, 
and assuming a Galactic-type extinction law (Fitzpatrick 1999)
and the starburst absorption curves of
Calzetti et al. (1994).
Most reddening values $E_{B-V}$ for Antennae 
clusters lie between 0.0 and 1.0, 
with a median value $\approx0.3$~mag, very similar to that
determined by Whitmore et al. (1999) from reddening-free
$Q$ parameters. 
The standard error in $\log\tau$ for an
individual cluster is $\sim0.3$--0.4, corresponding to a factor
of 2.0--2.5 in $\tau$, determined from a comparison with
spectroscopic ages for $>20$ Antennae clusters (FCW05;
R. Chandar et al. 2009, in preparation).
For most ages, the errors in $\log\tau$ are approximately symmetric, 
causing little if any bias in the age distribution. 
This is not true, however, for the very youngest clusters and those
with ages in the range $7.0 \la \log(\tau/{\rm yr}) \la 7.5$, where 
the optical emission from massive clusters is dominated by red 
supergiant (RSG) stars.
During the RSG phase, the color tracks of the stellar population 
models loop back on themselves, and the fitted ages become 
degenerate, with a tendency to avoid values just above $10^7$~yr 
and a tendency to prefer slightly higher values. 
This well-known effect accounts for the  
relatively-empty vertical stripes in Figures 1 and 2.  
As a result of this type of bias, there
may be small-scale features in the observed age distribution 
that are not present
in the real age distribution. We deal with this problem
(in \S3.3--3.5; see also FCW05) simply by binning on somewhat 
larger scales, $\Delta\log\tau\sim0.6$--0.8, thus
smoothing over any fine structure in the  
age distribution, whether real or artificial.
Almost all studies of cluster age and mass distributions
rely on
broadband optical measurements similar to those used here,
and thus inherently face a similar problem.
Future studies that include near-infrared photometry,
which provides more reliable estimates of ages in the range 
$7.0 \la \log(\tau/{\rm yr}) \la 7.5$, may be less affected
by these difficulties.

We now consider the uncertainties in our
mass estimates.
The random errors in the ages propagate into $1\sigma$ 
uncertainties 
of $\approx 0.3$ in $\log M$, or a factor of two in $M$.  
As we have already noted, the derived masses, but not the ages, of
the clusters depend on the IMF in the stellar population models. 
We have adopted the Salpeter (1955) IMF, mainly for ease of
comparison with our earlier studies and those of others.
If we had adopted the more modern 
%(and doubtless more realistic)
Kroupa (2001) IMF or Chabrier (2003) IMF, which flatten below
1~$M_{\odot}$, all the masses in this paper would be reduced by 
a nearly constant (age-independent) offset of 40\% (shown
graphically in CFW09).
Similarly, adopting a shorter (longer) distance to the Antennae
would systematically reduce (increase) the cluster masses, since
they are derived from luminosities.
It is important to note that none of these systematic 
uncertainties affect the {\em ratios} of cluster masses or the 
shape of the mass function presented in \S3.2, and hence they do not 
affect the primary results of this paper.

\subsection{Mass and Luminosity Functions}

Figures~1 and 2 contain much of the statistical 
information about the population of clusters in the Antennae,
embodied in the bivariate mass-age distribution $g(M, \tau)$
and its various projections, including the univariate mass,
age, and luminosity distributions.
In FCW05, we presented the age distribution
from this data set, and in this section,
we present new determinations of the mass and
luminosity functions.

The mass function can be obtained by projecting clusters
diagonally, along the stellar population tracks in Figure~1, 
or equivalently, horizontally in Figure~2.
Figure~3 shows the mass function
for Antennae clusters in the two age intervals
$10^6~{\rm yr} < \tau \le 10^7~{\rm yr}$ and
$10^7~{\rm yr} < \tau \le 10^8~{\rm yr}$, which were 
chosen in part to minimize the influence of errors in
the estimated ages during the RSG phase mentioned above.
We adopt a bin width of 0.4 in log~$M$ as a compromise
between uncertainties in the mass estimates and
adequate sampling of the mass function.
We have checked that our mass function is not
sensitive to the specific (constant or variable)
widths or locations of the bins, provided that they
are at least $\approx$0.2 wide in log~$M$.
The plotted range of masses for each age interval was 
chosen to lie above the stellar contamination limit,
except for the lowest mass bins, which contain
some  objects fainter than this limit (fewer than 50\%).
We find that over this mass-age domain,
the mass function declines monotonically
with no obvious breaks or bends.
It can be represented nicely by a pure power law, 
$dN/dM \propto M^{\beta}$, with $\beta=-2.14\pm0.03$ 
for clusters with $10^6~{\rm yr} < \tau \le 10^7~{\rm yr}$ 
and $M\ga 10^4~M_{\odot}$, and $\beta=-2.03\pm0.07$ for 
clusters with $10^7~{\rm yr} < \tau \le 10^8~{\rm yr}$ 
and $M\ga3\times10^4~M_{\odot}$, 
where the exponents and their 1$\sigma$ errors are based on 
least-square fits of the form 
$\log(dN/dM) = \beta \log M + {\rm const}$.
This power law extends at least up to $10^6 M_{\odot}$
and possibly up to $10^7 M_{\odot}$.
The true uncertainties in the exponents are larger than
the quoted formal errors, probably $\Delta\beta \approx 0.2$, 
based on our experience with different stellar population models, 
extinction laws, and bin sizes.

The mass function of the Antennae clusters presented here is in 
excellent agreement with the earlier one from ZF99, based on the same 
WFPC2 observations, and with a more recent one from Whitmore et al.
(2009), based on new observations with the ACS on {\em HST}.
However, it differs from the mass functions for the Antennae clusters 
presented by Fritze-von Alvensleben (1999) and Mengel et al. (2005), 
both of which exhibit substantial deviations from power laws.
Fritze-von Alvensleben found a lognormal-like mass function with
a peak or turnover at $\sim10^5~M_{\odot}$.
This feature, however, was close to the limit of the WF/PC1 
observations (with spherical aberration), and was subsequently 
shown to be an artifact of incompleteness by the deeper WFPC2 
observations (ZF99).
Mengel et al. (2005) found a mass function with a gradual bend at
$\sim 10^5 M_{\odot}$.
However, this is based on a sample of clusters with a wide range 
of ages and a constant luminosity limit, rather than one that tracks 
the fading of the clusters (as in the ZF99 and present studies).
The Mengel et al. selection criteria therefore include younger 
low-mass clusters, but artificially exclude older low-mass 
clusters, those that have faded below the fixed luminosity limit.
The bend in the mass function then reflects this underrepresentation
of low-mass clusters in the sample, not any physical process
involving the formation or disruption of the clusters.
We demonstrate this explicitly in Appendix A, where we construct
mass functions from our own sample with selection criteria like 
those adopted by Mengel et al.

It is also instructive to determine the luminosity function
of the Antennae clusters and its relationship to the mass
function.
While the mass function is usually of greater interest in
dynamical studies, the luminosity function can be derived from
observations in fewer bands and is available for the cluster
systems in more galaxies.
Figure~4 shows our determinations of the luminosity function of 
the Antennae clusters, with and without corrections for extinction.
These can be represented by pure power laws, 
$dN/dL \propto L^{\alpha}$, with $\alpha=-2.09\pm0.05$ 
(corrected for extinction) and $\alpha=-2.37\pm0.06$
(not corrected for extinction), for $L_V\geq3\times10^5L_{\odot}$.
While there may be minor deviations from these power laws, 
such as mild curvature, at the level $\Delta\alpha \la 0.3$, 
the reality of such features is doubtful, given that the true 
uncertainties on $\alpha$ are nearly as large.
We conclude that the extinction-corrected luminosity function
has the same or nearly the same power-law shape as the mass 
function, i.e., $\alpha\approx\beta\approx-2$.

Before we discuss the implications of this result, we note 
a contrary claim by Anders et al. (2007).
They find a peak in the $V$-band luminosity function of the 
Antennae clusters near $M_V\approx-8.5$.
This is based on a sample that excludes all clusters 
with uncertainties in the $U$-band magnitudes greater than $0.2$.
However, these clusters certainly exist and should be included
in any unbiased determination of the luminosity function, as
they are in the results presented here. 
When we apply the Anders et al. selection criteria to our own
sample, we reproduce their luminosity function, as discussed 
further in Appendix A.
Thus, we conclude that the peak they have found is an
artifact of their selection criteria. 

In a previous study of the Antennae clusters, we found some 
marginal evidence for a weak bend in the luminosity function 
near $M_V\approx-10.4$ (Whitmore et al. 1999).
This was based on luminosities without corrections for extinction.
There is a slight hint of mild curvature in the uncorrected and
corrected luminosity functions derived here, but we are skeptical
of this feature because it appears at about 
the same level as the true uncertainties in the exponent,
namely $\Delta\alpha \approx 0.2$.
A recent analysis of deeper observations taken with the ACS on $HST$ 
confirms the results presented here that the luminosity function of 
the Antennae clusters is well represented by a pure power law with 
$\alpha\approx-2$ (Whitmore et al. 2009).
Moreover, the luminosity functions of the star clusters in many other
galaxies can also be approximated by power laws with $\alpha\approx-2$ 
(Larsen 2002).

While the luminosity function is sometimes regarded as a proxy 
for the mass function, this need not be true for clusters with a 
wide range of ages and hence mass-to-light ratios.
For example, a (hypothetical) population of star clusters that all
form with the same mass at a constant rate and experience no disruption
will have a power-law luminosity function with $\alpha\approx-2$, very 
different from the assumed underlying delta-function mass 
function (Fall 2006).
Thus, the similarity of $\phi(L)$ and $\psi(M)$ we have found for the 
Antennae clusters should not be taken for granted.
Instead, it must reflect a basic property of the mass-age distribution
$g(M,\tau)$. 
In the special case that this distribution is separable, $g(M,\tau) 
\propto \psi(M) \chi(\tau)$, and the mass function is 
a power law, $\psi(M) \propto M^{\beta}$, the age variable can be 
integrated out and the luminosity function will also be a power law, 
$\phi(L) \propto L^{\alpha}$, with the same exponent, $\alpha = \beta$ 
(Fall 2006).
This is true regardless of the specific form of $\chi(\tau)$.
Therefore, the fact that the observed mass and luminosity functions 
of the Antennae clusters are similar power laws is indirect  
support for the decomposition of $g(M, \tau)$ given in 
equation~(1).
Indeed, the assumption that masses and ages are independent of one 
another is the basis of Model~3, which, as we show in $\S3.5$,
provides a much better fit to the observations than Models 1 and
2.

\subsection{Comparison with Model 1: Sudden Mass-Dependent Disruption}

In this and the next two subsections, we compare predictions from the
three disruption models developed in \S2 with observations
of the Antennae clusters. 
The simplest way to do this is in terms of the
{\em averages} of $g(M,\tau)$ over several adjacent intervals of 
mass and age. 
These averages or projections are given by the expressions

\begin{equation}
\bar{g}(M) \equiv \frac{1}{(\tau_2 - \tau_1)}
\int_{\tau_1}^{\tau_2} g(M, \tau) d\tau,
\end{equation}

\begin{equation}
\bar{g}(\tau) \equiv \frac{1}{(M_2 - M_1)}
\int_{M_1}^{M_2} g(M, \tau) dM.
\end{equation}

\noindent
An alternative approach would be to compare the models and observations 
in the $M$--$\tau$ plane, using standard statistical methods
(e.g., minimum $\chi^2$ or maximum likelihood)
to derive a single goodness-of-fit measure.
This however, has the disadvantage that any systematic errors,
like those from estimating ages in the RSG phase
as discussed in \S3.1, cause artificial non-random deviations
from the models, 
for which it is not possible to assign meaningful
$\chi^2$ values.
Moreover, it turns out to be fairly obvious from $\bar{g}(M)$ and
$\bar{g}(\tau)$ which models match the 
observations and which do not.
We present analytical formulae for  
$\bar{g}(M)$ and $\bar{g}(\tau)$ 
in Appendix~B for the three models
discussed in \S2.
These are our main tools
for interpreting the observations in terms of
formation and disruption processes.

We first consider
Model~1, which assumes that clusters are disrupted suddenly
at a mass-dependent age $\tau_d(M) = \tau_*(M/M_*)^{k}$.
It is instructive first to display the predictions of
this model in the $M$--$\tau$ plane.
Figure~5 shows a Monte Carlo realization of a cluster 
population generated by Model 1 with $\beta = -2$, 
$\tau_*=5\times10^7$~yr, and $k=0.62$.
It has been suggested that this value of $k$ may be 
ubiquitous among the cluster systems of different galaxies 
(Boutloukos \& Lamers 2003).
In this model, lower mass clusters are destroyed sooner 
than higher mass clusters, resulting in a sharp, diagonal 
line beyond which no clusters survive.
The index $k$ regulates the slope of this
line  (proportional to $1/k$), while
$\tau_*$ controls its intercept.   
Model~1 predicts an evacuated, triangular region to the
right of the disruption line (marked D in Fig. 5).
However, no such feature appears in Figure 2, the observed 
$M$--$\tau$ plane for the Antennae clusters.
Thus, we can immediately rule out Model 1, at least
for the mass-age domain studied here.

Despite this failure of Model 1, it is still instructive 
to compare the predicted and observed 
$\bar{g}(M)$ and $\bar{g}(\tau)$ distributions as
a prelude to tests of the more realistic Models~2 and 3.
Figures~6 and 7 show these comparisons for the same parameters 
adopted in Figure 5.
Evidently, the {\em shapes} of the predicted and observed 
$\bar{g}(M)$ distributions match for sufficiently large 
$\tau_*$ (i.e. $\tau_* \gea5\times10^7$~yr), ensuring that
the predicted curvature occurs below the observed mass range.
However, the {\em amplitudes} cannot be matched; the predicted 
amplitude is constant with age because, in the model, there 
is no evolution until clusters are suddenly disrupted, whereas 
the observed amplitude decreases substantially with age.
The opposite problem arises with the $\bar{g}(\tau)$ 
distribution.
In this case, the predicted and observed {\em amplitudes} 
agree (a direct consequence of the adopted initial mass 
function, $\psi_0(M) \propto M^{-2}$), but the {\em shapes} 
are quite different;
the observations decline approximately like a power law in 
age, while the predictions remain flat for a while and then 
drop rapidly near the disruption time $\tau_d$ for clusters 
with the relevant mass ($M_1 < M < M_2$).
This behavior is generic: $\bar{g}(\tau)$ is predicted to
have a sharp bend at $\tau_d$, but none is observed.
Thus, we confirm our earlier conclusion that Model 1 does 
not describe the population of massive young clusters in the 
Antennae.

\subsection{Comparison with Model 2: Gradual Mass-Dependent Disruption}

Model~2 retains the power-law dependence of the disruption
timescale on mass but smooths out the discontinuous evolution
postulated by Model~1.
As a result of this smoothness, one might guess that Model 2
would fare better against the observations than Model 1, but
this turns out not to be true, as we now demonstrate.
In Figures~8, 9, and 10, we compare the predicted $\bar{g}(M)$ 
from Model 2 for $k = 0$, 0.6, and 1.0, respectively, with 
our Antennae observations.
In the limit $k = 0$, the disruption rate is independent 
of mass.
In this case (Fig. 8), the predicted and observed $\bar{g}(M)$ 
match in shape for all values of $\tau_*$ but in amplitude only 
for $\tau_*\approx1.4\times10^7$~yr.
However, with this value of $\tau_*$ (or any other), the predicted 
$\bar{g}(\tau)$  fails to reproduce the observed one, as shown in 
Figure~11.
The reason for this is that, with the disruption law postulated by
Model 2 (eqs. [8] and [9]), there is always a bend near $\tau_d$ 
in $M(\tau)$ (eq. [10]) and hence in $\bar{g}(\tau)$ (eqs. [B6] and
[B7]), whereas the observations show no such feature.
In the mass-dependent cases, $k=0.6$ (Fig.~9) and $k=1.0$ (Fig.~10), 
the predicted 
$\bar{g}(M)$ matches the observed one in shape for large enough
$\tau_*$, so that the predicted curvature occurs below the 
observed mass range, but it never matches in amplitude for
clusters both younger and older than $10^7$~yr.
These are essentially the same problems that afflicted Model~1,
with relatively minor differences resulting from the different
disruption laws (one sudden, one gradual).
We conclude, therefore, that Model 2 also fails to describe the
Antennae clusters in the mass-age domain studied here.

This leads us to a more general consideration of mass-dependent
disruption.
%As Figures 3 and 6--13 show, 
We have shown that the observed mass function of
star clusters in the Antennae has the 
same shape but different amplitude for different ages, while the 
observed age distribution has the same shape but different amplitude 
for different masses. 
These facts cannot be explained by mass-dependent disruption.
In particular, if low-mass clusters were disrupted faster than
high-mass clusters, the mass function would become flatter with
increasing age; and conversely, if disruption had the opposite
dependence on mass, the mass function would become steeper.
Neither effect is observed in the mass function of the
Antennae clusters.
Furthermore, there are no discernible features in the 
observed mass and age distributions, as would be expected 
if the rate of disruption changed at a characteristic mass or age.
We emphasize that these conclusions are not affected by
the formation history of the clusters (i.e., whether they form
at a constant or variable rate), provided only that the shape 
of their initial mass function is invariant.
In this case, variations in the formation rate would affect 
only the amplitude of the mass function, but not its shape, 
at different ages.\footnote{
The fact that the shape of the mass function is not altered
by a variable rate of cluster formation also follows 
from equations~(3) and (4), where $C(\tau)$ appears only as 
a prefactor in $g(M,\tau)$.}
%This is clear physically and can also be demonstrated
%mathematically from equations~(3) and (4).
Thus, we conclude, quite generally, that disruption of the
massive young clusters in the Antennae has little or no 
dependence on mass.

\subsection{Comparison with Model 3: Gradual Mass-Independent Disruption}

Model~3, with mass-{\em in}dependent disruption, was designed to remedy 
the problems encountered in fitting Models~1 and 2 to the Antennae data.
As we noted in Section 2, the disruption of the population of clusters
is gradual in this model, although the disruption of individual clusters
may be gradual or sudden. 
This model also has fewer adjustable parameters ($C$, $\beta$, $\gamma$)
than the other models ($C$, $\beta$, $k$, $\tau_*$).
In Model~3, the averages over $g(M, \tau)$ are pure power laws:
$\bar{g}(M) \propto M^{\beta}$ and $\bar{g}(\tau) \propto \tau^{\gamma}$.
As Figures 12 and 13 show, the predicted and observed $\bar{g}(M)$ and
$\bar{g}(\tau)$ agree nicely, both in shape and amplitude, for 
$\beta = -2.0$ and $\gamma = - 1.0$.
These results corroborate those from our earlier studies of the 
Antennae clusters (ZF99, FCW05, WCF07). 
Model 3 provides an excellent description of the population of
massive young clusters in the Antennae. 
We note that the domain of validity of this result---the region of
the $M$--$\tau$ plane above the stellar contamination limit---covers 
a relatively wide range of masses for very young clusters but becomes 
progressively narrower with increasing age, as shown in Figure 2 and 
approximated by equation (2).
Thus, we regard Model 3 as a robust description of the 
observations for $\tau \lea 10^8$~yr and more akin to an 
extrapolation for $\tau \gea 10^8$~yr.

\section{INTERPRETATION AND IMPLICATIONS}

We showed in the previous section that the joint distribution of masses
and ages of massive young clusters in the Antennae can be approximated
by the simple formula $g(M,\tau) \propto M^{-2}\tau^{-1}$.  
This means that the shape of the mass function $\psi(M) \propto M^{-2}$ 
is independent of age and that the shape of the age distribution 
$\chi(\tau) \propto \tau^{-1}$ is independent  of mass 
(at least over the observed domain of $M$ and $\tau$).
We now seek to understand these results in terms of the physical 
processes that operated during the formation and early evolution 
of the clusters. 
We also consider the long-term future evolution of the young 
clusters and whether their mass function will eventually resemble
that of old globular clusters.
The picture presented here is an extension of the one proposed
by Harris \& Pudritz (1994) and Elmegreen \& Efremov (1997) and
developed in some of our earlier papers (ZF99; Fall \& Zhang 2001;
FCW05).

\subsection{Formation of Star Clusters from Molecular Clouds}

The mass functions of young star clusters must depend in some way
on the mass functions of the molecular clouds in which
they formed.
Blitz et al. (2007) review CO observations of molecular
clouds in several galaxies in the Local Group: the Milky Way,
LMC, SMC, M31, M33, and IC10.  The mass functions of the clouds
can be represented by power laws, $dN/dM \propto M^{\beta}$,
with $\beta\approx-1.7$ in most cases.  
The exceptions are M33, which has $\beta\approx-2.5$, 
and the SMC, which has too few clouds
to estimate $\beta$.
The range of masses covered by these observations varies
substantially, from a
$\mbox{few}\times10^{5}~M_{\odot} \lea M \lea 10^6~M_{\odot}$
for M33 to
$10^3~M_{\odot} \lea M \lea \mbox{few}\times10^6~M_{\odot}$
for the Milky Way.
For complexes of molecular clouds in the Antennae, 
Wilson et al. (2003) find $\beta \approx-1.4$
(for $M \gea10^7~M_{\odot}$).
While these estimates of $\beta$ are suggestive,
it is worth noting that they pertain to clouds with
relatively low column density thresholds (compared
with the regions of most intense star formation)
and, in the case of the Antennae, may also
suffer from blending as a result of the larger distance and hence
lower spatial resolution.

It is also interesting at this stage to recall some results for
protoclusters in other galaxies. Studies of
HII regions and OB associations in the Milky Way and many nearby galaxies
show that their H$\alpha$ and Lyc luminosity functions have power-law
form, $dN/dL \propto L^{\alpha}$, with exponents near $\alpha = -2$ 
(Kennicutt et al. 1989; McKee \& Williams 1997). 
There appear to be small differences
among the exponents for different galaxies and in some cases a
high-luminosity turnover or cutoff. As we have noted before,
luminosity and mass functions in general are not the same, due to the
spread in ages and fading of the clusters. 
However, for HII regions
and OB associations, the age spread is relatively narrow ($\tau \la
10^7$~yr), the $M/L$ variations are modest, and luminosity functions 
become effective proxies for mass functions. 
Dowell et al. (2008) have derived the mass
functions of young clusters ($\tau \la 2 \times 10^7$~yr) in several
nearby galaxies based on broadband multi-color photometry (by a method
generally similar to ours but using ground-based rather than $HST$ data). 
For $M \ga {\rm few}\times 10^4 M_{\odot}$, 
they find $dN/dM \propto M^{\beta}$, with $\beta = -1.9$ for their
combined sample of irregular galaxies and $\beta = -1.8$ for their
combined sample of spiral galaxies. 
We find $\beta\approx-1.8$ for the young clusters in the LMC (CFW09).

It is remarkable that all these mass functions are power
laws with nearly the same exponent, $\beta \approx -2$. There may be
small but real differences between the mass functions of molecular
clouds and young star clusters in general and/or between the mass
functions of either clouds or clusters in different galaxies, at the
level $\Delta\beta \approx 0.2$--0.6. 
The significance of these
differences is low, typically 1--3~$\sigma$, based on the
statistical and likely systematic uncertainties in $\beta$.
We also note that the different mass
functions were determined by different researchers using
different methods and that they pertain to different density
thresholds and mass ranges.
Thus it is possible that they are all basically the same, 
or more conservatively, that any differences between them are
barely detectable. 
This means that the average efficiency of star formation in 
protoclusters---the ratio of the final stellar mass to the 
initial interstellar mass---is
approximately independent of the mass of the
protoclusters.\footnote
{In principle, the relation between the mass functions of 
the clouds and clusters also depends on the lifetimes of the
clouds, although these are expected to have at most a
weak dependence on mass (Matzner \& McKee 2000).}
If the efficiency varies at all, 
it decreases with increasing mass, since the mass functions
of molecular clouds appear to be slightly flatter than the 
mass functions of young clusters.  
On average, low-mass protoclusters make stars just
about as efficiently as high-mass protoclusters. 
This is a simple but important empirical result,
which should help to guide theories of star and
cluster formation.
Indeed, it probably reflects a balance
between star formation and stellar feedback within
protoclusters (Elmegreen \& Efremov 1997; Krumholz
et al. 2006).

A complete understanding of the mass functions of young star clusters
clearly requires an understanding of the mass functions of their
progenitor molecular clouds.
The ISM is often described as a turbulent medium in which
the density and velocity fluctuations on each spatial scale
mimic those on other scales in a self-similar hierarchy
(see the reviews by Elmegreen \& Scalo 2004 and McKee \&
Ostriker 2007).
Given this picture, it is not surprising that the mass
function of molecular clouds is a power law.
The important question is what physical processes determine
the exponent $\beta$.  
Fleck (1996) has derived
$\beta=-2$ for relatively diffuse, non-self-gravitating
clouds based on a simple analytical model of compressible
turbulence in the ISM.  
Wada et al. (2000) find $\beta=-1.7$ in their 2D hydrodynamic 
simulations of ISM turbulence without stellar feedback and
slightly steeper mass functions when feedback is included
(see also  V{\'a}zquez-Semadeni et al. 1997).
Elmegreen (2002) finds $\beta\approx-2$ in his static models
of a fractal ISM, and slightly steeper (shallower)
mass functions for clouds with higher (lower) density
thresholds.
Despite these results, it is not yet entirely clear why the 
mass function of real molecular clouds has $\beta \approx-2$ 
rather than $\beta\approx-3$ or $\beta\approx-1$ (say).
Explaining the physical origin of this exponent remains an
important theoretical challenge.

\subsection{Early Disruption: Processes and Timescales}

Interstellar material is removed from protoclusters by feedback from
massive stars (photoionization, radiation pressure, stellar 
winds, and supernovae). 
Observationally, the dividing line between protoclusters
that contain interstellar material and those that do not is roughly
$\tau \sim 10^7$~yr, based on the coincidence of molecular clouds and
HII regions with the locations of clusters of different ages
(Blitz et al. 2007).
%(refs?). 
This timescale also makes sense from a physical point of
view; it is a few multiples of the lifetimes of the most massive stars
and about an order of magnitude longer than the dynamical or crossing
times within the clusters. 
An important consequence of this early,
rapid mass loss is that many of the protoclusters cannot remain
gravitationally bound and therefore begin to dissolve almost as
soon as they form (Hills 1980), 
a process often called ``infant mortality.''
In reality, the removal of ISM by stellar feedback will unbind
some protoclusters completely and others only partially, leaving
behind less-massive, but bound remnants. 
This process obviously affects the age distribution and hence 
the {\em amplitude} of the mass function of young clusters.
But does it also affect the {\em shape} of the mass function?

It has been proposed that ISM removal disrupts
a higher proportion of low-mass clusters than high-mass
clusters, and that this introduces a bend or flattening in 
the mass function at $\sim 10^5 M_{\odot}$
(Kroupa \& Boily 2002; Baumgardt et al. 2008; Parmentier et al. 2008).
We believe these claims are based on unrealistic
models of protoclusters (see below). 
However, it is an easy matter to check observationally whether or not
they are correct, by comparing the mass functions of
Antennae clusters in the two age bins divided at $\tau = 10^7$~yr
(Figure~3). 
These show no hint of a bend; the power-law form
of the mass function in the older bin appears identical to that in the
younger bin, as we have already noted. 
If infant mortality imprints any feature in the mass
function at all, it must be well below $10^5 M_{\odot}$ and probably below
$10^4 M_{\odot}$ (to avoid any hint of curvature).
We obtain even stronger constraints on mass-dependent
disruption in our recent study of the clusters in the LMC (CFW09). 

The shape of the mass function of young clusters depends on how
effectively stellar feedback removes the ISM from protoclusters 
of different masses.
This in turn depends on the fraction of the input energy that is
radiated away from the protoclusters as a function of their mass.
Most of the internal kinetic energy of molecular clouds is in the 
form of supersonic turbulence, which is dissipated in strong shocks 
on the dynamical timescales of the clouds (McKee \& Ostriker 2007 
and references therein).
This important energy sink is not included in the models proposed
by Kroupa \& Boily (2002), Baumgardt et al. (2008), and Parmentier 
et al. (2008).
Thus, it is likely that they have overestimated the energy available 
for disrupting protoclusters, possibly by large factors.
In the limit of no radiative losses, the feedback is 
energy-driven, whereas in the opposite limit of maximum losses, 
it is momentum-driven.
The second regime is likely to be more realistic than the 
first, and further analysis is needed to determine whether such
feedback would imprint any features on the mass function of young 
clusters. Preliminary indications are that it would not
(S. M. Fall et al. 2009, in preparation).

The preceding discussion of infant mortality presumes that the 
molecular clouds in which star clusters formed were gravitationally
bound for at least some time before the onset of stellar feedback.
This picture, however, is now debated (see the reviews by
Elmegreen \& Scalo 2004 and McKee \& Ostriker 2007).
It is possible instead that some of the sites of star and
cluster formation are merely transient molecular concentrations
created by convergent flows in the interstellar turbulence. 
The protoclusters produced at these sites might be unbound from the
beginning, irrespective of any stellar feedback that occurs within
them. This cannot be the whole story, of course, because at least a
small fraction of clusters---those that survive for many of their own
crossing times---must have formed as gravitationally bound objects. 
The revised picture in which some protoclusters are initially
bound while others are unbound has not yet been studied in enough
detail to make firm predictions about the shape of the resulting mass
function of the surviving clusters. 
In a scale-free (fractal) ISM, however,
there is no immediate reason to suspect that the fractions of bound 
and unbound clusters would change abruptly at a preferred mass scale 
or would even depend on mass at all.  

The young clusters will continue to lose mass through stellar winds
and other ejecta even after they have expelled all their nascent
interstellar material. 
Stellar evolution alone depletes the masses of
star clusters by $\sim$40\% over a period of a ${\rm few} \times
10^8$~yr, although most of this mass is lost in the first ${\rm
few}\times10^7$~yr, the exact fraction and timescale depending on the
shape of the stellar IMF. 
This process can unbind clusters in a tidal field if they are 
already weakly bound, with concentration parameters 
[$c \equiv \log (r_t/r_c)$] in
the range $c \la 0.7$ (corresponding to dimensionless central
potentials in the range $W_0 \lea3$; Chernoff \& Weinberg 1990;
Fukushige \& Heggie 1995; Takahashi \& Portegies Zwart 2000). 
As a result of the prior removal of interstellar material, 
we expect many of the surviving clusters to be only 
weakly bound and thus vulnerable to disruption by 
subsequent stellar mass loss. 
If the concentration parameters of the clusters are 
uncorrelated with their masses (the
simplest possibility), a large fraction of them could be disrupted in
the period $10^7~\mbox{yr} \la \tau \la 10^8$~yr, without altering the
power-law shape of the mass function, consistent with our observations
of the Antennae and LMC clusters. 
In contrast, Vesperini \& Zepf (2003) have
argued that this process would produce a ``bell-shaped'' mass
function. 
They base this claim on an assumed strong correlation
between concentration and mass for young clusters, a hypothesis for
which there is no physical explanation or observational evidence. 
The old globular clusters in the Milky Way do have such a 
correlation, but this is almost certainly a product of internal 
dynamical evolution over a Hubble time, including core collapse, 
tidal heating, and so forth. 
In any case, the similar shapes of the mass functions of clusters
younger and older than $10^7$~yr in both the Antennae and the LMC 
tell heavily against this suggestion.

Another mechanism for disrupting star clusters is tidal disturbances by 
passing molecular clouds (Spitzer 1958). 
We follow the comprehensive treatment by Binney \& Tremaine (2008, Sec 8.2) 
and distinguish two regimes: catastrophic, in which clusters are disrupted 
suddenly by a single strong encounter, and diffusive, in which clusters are 
disrupted gradually by a sequence of weak encounters.  
The disruption timescales in these regimes (up to proportionality) are
\begin{equation}
t_{\rm d} \propto \frac{\rho_{\rm h}^{1/2}}{M_{\rm p} n_{\rm p}}    
\,\,\,\,\,\,\,\,\,\,\,\,\,\,\,~~~~~~~~~~ {\rm (catastrophic~ regime)},
\end{equation}
\begin{equation}
t_{\rm d} \propto \frac{\sigma_{\rm rel} r_{\rm hp}^2 \rho_{\rm h}}
{M_{\rm p}^2 n_{\rm p}}                                                                          
\,\,\,\,\,\,\,\,\,\,\,\,\,\,\,~~~~~~~~~~~ {\rm (diffusive~ regime)}.
\end{equation}
Here, $\rho_{\rm h} \equiv 3 M /(8 \pi r_{\rm h}^3)$ is the mean density 
within the half-mass radius $r_{\rm h}$ of the clusters, $M_{\rm p}$, $r_{\rm hp}$, 
and $n_{\rm p}$ are the mass, half-mass radius, and mean number density of the 
molecular clouds (the perturbers), respectively, and $\sigma_{\rm rel}$ is the 
dispersion of the relative velocities between the clusters and clouds.
Equations (16) and (17) display two important consequences of
tidal disruption: 
(1) In both regimes, $t_{\rm d}$ depends on the masses $M$ of the clusters 
only through their mean internal densities $\rho_{\rm h}$; 
(2) In the catastrophic regime, $t_{\rm d}$ depends on the properties of 
the molecular clouds only through their mean smoothed-out 
density, $\bar{\rho}_{\rm p} \equiv M_{\rm p} n_{\rm p}$.
For open clusters in the solar neighborhood, Binney \& Tremaine (2008) 
estimate $t_{\rm d} \sim 3 \times 10^8$~yr 
(in the catastrophic regime). 
This might provide a rough indication of what to expect in the Antennae galaxies, 
since $\rho_{\rm h}$ and $\bar{\rho}_{\rm p}$ might also be similar to their 
local values.  
If so, tidal disturbances would be effective in disrupting clusters older 
than $\sim10^8$~yr in the Antennae.
Despite the large uncertainty in the timescale for this process, we do not expect 
it to alter the shape of the mass function, since the mean internal densities of 
the clusters are fixed mainly by the galactic tidal field and should therefore 
be approximately the same for clusters of different masses. 

We argued in the Introduction that the age distribution $\chi(\tau)$
primarily reflects the disruption history rather than the formation
history of the clusters. The disruption history in turn is likely
dominated by different physical processes in different intervals of
age. 
Based on our previous discussion, the following sequence seems
plausible: (1) disruption by removal of
ISM, $\tau \la 10^7$~yr; (2) disruption by stellar mass loss, $10^7
{\rm yr} \la \tau \la 10^8 {\rm yr}$; (3) disruption by tidal
disturbances, $\tau \ga 10^8$~yr. 
It also seems likely that the
rate of disruption and hence the shape of the age distribution varies
from one of these regimes to another, with some sort of features at
the transitions between them. 
However, this combination of disruption processes involves too many
unknowns to predict how prominent such features would be.
As we have already noted, the uncertainties in the ages of the clusters 
preclude the measurement of fine structure in the age distribution, 
especially near the likely transition between disruption dominated 
by ISM removal and by stellar mass loss, at $\tau \sim 10^7$~yr.
Moreover, the tail of $\chi(\tau)$ potentially dominated by tidal
disturbances, at $\tau \ga 10^8$~yr, is based on observations near
the limit of our sample, where the accessible range of masses is 
relatively small.
Plausible variations in the formation rate could also make
the age distribution slightly flatter or slightly steeper. 
Thus, it is likely that our
power-law model, $\chi(\tau) \propto \tau^{-1}$, is a simple
approximation to a complex situation involving several different
physical processes rather than an exact description of a single
process. In any case, since the shape of the mass function is
preserved by each of these processes, it will also be preserved by any
combination of them.

\subsection{Late Disruption of Clusters: Evaporation}

As we already noted, a primary motivation for the present study has been
to understand the formation and early evolution of star clusters in an
environment similar to that during the early phases of the
hierarchical assembly of galaxies, when the clusters that are now
classified as old globular clusters were originally produced. Thus, we
are led to speculate on the long-term future evolution of the young
clusters in the Antennae, those that survive the first $\sim 10^8$~yr,
over the next $\sim 10^{10}$~yr or so. A major issue here is that the
mass functions of the young and old clusters are very different, and
this immediately raises the question whether the former would evolve
by long-term disruptive processes into the latter.

The upper panel of Figure 14 shows the mass functions of the young
clusters in the Antennae and the LMC (both samples
restricted to $\tau \la 10^8$~yr). 
Here, we have plotted $dN/d\log{M}$
rather than $dN/dM$,
as in our previous figures. 
The mass function for the Antennae clusters is taken
from the present paper, while that for the LMC clusters is taken from
our companion paper (CFW09), with a vertical shift to allow for the
different sizes of the two populations. Both functions have
the same power-law shape, although they cover different ranges of mass
(higher in the Antennae, lower in the LMC). The upper smooth curves in
both panels of Figure 14 are the same Schechter function, $\psi_0(M)
\propto M^{\beta} \exp (-M/M_c)$, with the parameters $\beta = -2$ and
$M_c = 2 \times 10^6 M_{\odot}$. We have appended the subscript 0 to
$\psi$ to indicate that we will adopt this as the ``initial'' mass
function in the calculation that follows. 
Evidently, we can represent the mass functions of {\it young\/} clusters 
over the observed range of masses equally well by a pure power law
or a Schechter function with a sufficiently large cutoff $M_c$. In the
present context, the Schechter function is preferable because it
provides a better fit to the high-mass end of the mass function of
{\it old\/} globular clusters, shown in the lower panel of Figure~14
(see also Burkert \& Smith 2000; Jord{\'a}n et al. 2007).

The most important long-term disruptive process operating on clusters
is the gradual escape of stars driven by internal two-body relaxation
(``evaporation''). 
This process depletes the mass of each cluster approximately linearly
with time, $M(t) \approx M_0 - \mu_{ev}t$, at a rate $\mu_{ev}$ that
depends primarily on the mean internal volume or surface density of
the cluster ($\mu_{ev} \propto \rho_t^{1/2}$ with $\rho_t \propto M/r_t^3$
for standard evaporation, $\mu_{ev} \propto \Sigma_t^{3/4}$ with
$\Sigma_t \propto M/r_t^2$ for retarded evaporation; see Fig.~6 and 
eq.~(12) of Baumgardt \& Makino 2003 for the linearity of $M(t)$ and 
McLaughlin \& Fall 2008 for the dependence of $\mu_{ev}$ on
$\rho_t$ and $\Sigma_t$).
The mass function of clusters with the same internal density at an
age $\tau$ is then related to the initial mass function by the simple 
formula
\begin{eqnarray}
\psi(M, \tau) &\approx &\psi_0(M + \mu_{ev}\tau)\\\nonumber
&\propto &(M + \mu_{ev}\tau)^{\beta}\exp [-(M + \mu_{ev}\tau) /M_c] 
\end{eqnarray}
(Fall \& Zhang 2001; Jord{\'a}n et al. 2007). 
For a population of clusters with different internal densities, 
and hence evaporation rates, the mass function is a superposition 
of terms like the one above, each with its own value of $\mu_{ev}$. 
Here, we approximate this superposition by a single term with a 
typical value of $\mu_{ev}$ for the whole population, which we have
checked is sufficiently accurate for our present, illustrative 
purposes (see McLaughlin \& Fall 2008 for a comprehensive 
discussion of these issues).

Two other long-term
disruptive processes that are sometimes invoked are gravitational
shocks (during passages through a galactic disk or near a galactic
bulge) and dynamical friction. Gravitational shocks may be important
for some massive clusters, but they are negligible compared with
evaporation for low-mass clusters, in particular, those with masses
near and below the turnover in the mass function i.e., $M \la 10^5~
M_{\odot}$ (Fall \& Zhang 2001). 
Furthermore, gravitational shocks alone cannot change the
shape of the mass function.\footnote{In the 
notation of \S2,
with the disruption timescale written in the form $\tau_d(M)
\propto M^k$, evaporation corresponds to $k = 1$ and shocks to $k =
0$.} Dynamical friction is only important in a massive galaxy, such as
the eventual remnant of the merging Antennae, for a few very massive
clusters located very near the galactic center and can safely be
neglected for all the other clusters in a first 
approximation (Fall \& Zhang 2001; Binney \& Tremaine 2008). 
Tidal encounters with molecular clouds can also be neglected 
once the clusters are
dispersed out of the disks of the merging galaxies and into the
spheroid or halo of the remnant galaxy (at $\tau \sim 10^9$~yr or
earlier).

The lower panel of Figure 14 shows the mass functions of the old
globular clusters in the Milky Way and the Sombrero galaxy, again
plotted as $dN/d\log{M}$. Both mass functions are derived from the
corresponding luminosity functions, based on data from Harris (1996,
as updated at http://www.mcmaster.ca/~$\tilde{}$harris/mwgc.dat) and Spitler
et al. (2006), respectively, with an assumed mass-to-light ratio
$M/L_V = 1.5 M_{\odot}/L_{\odot}$, a typical value
obtained in dynamical studies (McLaughlin 2000). Again, the mass
functions were shifted vertically to allow for the different
sizes of the populations; and again both functions have similar
shapes, although they cover somewhat different ranges of mass. With
increasing mass, these functions first rise approximately linearly
($dN/d\log{M} \propto M$, corresponding to $dN/dM \approx {\rm
const}$), reach a peak at $M_p \approx (1-2)\times 10^5~M_{\odot}$,
and then decline steeply.
The five solid curves in
each panel of Figure 14 are the predictions of the simple evaporation
model, equation (18) with $\mu_{ev} = 2 \times 10^{-5}M_{\odot}~ {\rm
yr}^{-1}$ and $\tau = 0$, 1.5, 3, 6, and 12~Gyr.\footnote{The peak 
mass of $M\psi(M)$ increases approximately linearly with age: 
$M_p \approx \mu_{ev} \tau \approx 2 \times 10^5 
(\tau/10^{10}{\rm yr}) M_{\odot}$ (valid for $M_p \ll M_c$). 
The line specified by this equation lies well below the luminosity 
limit that defines our sample of Antennae clusters in the mass-age 
plane, i.e., below the solid curve in Fig. 2. Thus, we would not 
expect to see any evidence for such a feature in the mass functions 
of these clusters, consistent with the power laws shown in Fig. 3.}
This value of
$\mu_{ev}$ corresponds to an escape rate of 5\%--10\% per relaxation
time, as indicated by various Fokker-Planck, Monte-Carlo, and $N$-body
simulations of the evaporation of tidally limited clusters, when
combined with the typical observed internal densities of the globular
clusters in the Milky Way and Sombrero galaxies (Chandar et al. 2007;
McLaughlin \& Fall 2008). Thus, the horizontal position of the 
models in Figure 14 is fixed to within $\pm 0.2$ or less in $\log{M}$. 
The vertical position of the models is fixed by the high-mass end of 
the observed mass function, where evolution can be neglected.

Evidently, the model $\psi(M, \tau)$ at $\tau = 12$~Gyr coincides
remarkably well with the observed mass function of old globular
clusters over the full range of masses, from $\sim 10^4 M_{\odot}$ 
to above $\sim 10^6 M_{\odot}$. This is a very simple and gratifying
result: starting with a population of massive young clusters like 
those in the Antennae and the LMC, and then letting evaporation take 
its toll for about 12 Gyr, leads inevitably to a population of old 
globular clusters like
those in the Milky Way and the Sombrero galaxy. No other ingredients
or mechanisms are needed. This conclusion contrasts sharply with 
some recent suggestions that the turnover or bend at 
$M \sim 10^5~M_{\odot}$ in the evolved mass function of old globular
clusters is simply a relic of a similar feature in the mass function 
of young clusters (see, for example, Parmentier et al. 2008).  
There are two problems with this explanation: 
(1) The postulated bend in the ``initial'' mass function conflicts 
directly with observations, as we have shown here for the Antennae 
clusters (and was shown earlier by ZF99) and in a companion paper
for the LMC clusters (CFW09). (2) The claimed mechanisms for producing
a bend in the mass functions of young clusters are either physically 
unrealistic (neglecting radiative dissipation of turbulence), or 
unjustified (assuming a strong correlation between mass and 
concentration). 
We note that the model with a power-law initial mass function (i.e., 
without a bend) is both simpler and fully consistent with all 
relevant observations of old 
globular clusters, including the dependence of their evolved mass 
function on their internal densities and galactocentric positions 
(Chandar et al. 2007; McLaughlin \& Fall 2008).\footnote{The 
present orbital eccentricities
of old globular clusters are sometimes invoked as a constraint on the
shape of their initial mass function (Fall \& Zhang 2001; Vesperini
et al. 2003; Baumgardt et al. 2008).  This constraint, however, is
extremely weak in practice, because
it is based on the simplifying assumption that the galactic potential
in which the clusters orbit is spherical and static, an assumption 
that is not consistent with the hierarchical formation and evolution 
of galaxies (see Fall \& Zhang 2001; Chandar et al. 2007; McLaughlin 
\& Fall 2008 for further discussion of this point).}

\section{SUMMARY AND CONCLUSIONS}

In this paper, we have derived analytical formulae for
the bivariate mass-age distribution $g(M,\tau)$ for three
idealized models of the formation and disruption of star
clusters.
We have also derived formulae for the corresponding averages
of $g(M, \tau)$ over finite intervals of $M$ and $\tau$,
denoted by $\bar{g}(\tau)$ and $\bar{g}(M)$ respectively.
In all three models considered here, clusters are assumed to
form with a power-law initial mass function at a constant
rate.
In the first two models, proposed by Boutloukos \& Lamers
(2003), clusters are disrupted on a timescale that depends on
their masses as $\tau_d(M) \propto M^k$ with $k > 0$, either
suddenly (Model~1) or gradually (Model~2).
In the third model, clusters are disrupted (suddenly or
gradually) at a fractional
rate independent of their masses, as indicated by
our earlier studies of the Antennae clusters (ZF99, FCW05, WCF07).
Model~3 has the remarkably simple bivariate distribution
$g(M, \tau) \propto M^{\beta}\tau^{\gamma}$.
The corresponding luminosity, mass, and age distributions
are all pure power laws: $\phi(L) \propto L^{\alpha}$,
$\psi(M) \propto M^{\beta}$, $\chi(\tau) \propto
\tau^{\gamma}$.

We have compared Models 1, 2, and 3 with the empirical
distributions of luminosities, masses, and ages for a
large sample of star clusters in the Antennae galaxies.
These are based on our \textit{UBVI}H$\alpha$ photometry
of point-like sources in images taken with the WFPC2 on
{\it HST\/} and comparisons with stellar population models
to estimate $M$ and $\tau$ for each source (after correcting
for interstellar reddening).
To distinguish clusters of stars from individual
stars, we have restricted our sample for analysis to objects
brighter than $M_V = -9$, corresponding to $L > 3 \times
10^5 L_{\odot}$.
This limit defines the domain of validity of the models
in the $L$-$\tau$ and $M$--$\tau$ planes, the regions
above the solid curves in Figures~1 and 2 respectively,
corresponding approximately to $\tau \la 10^7
(M/10^4 M_{\odot})^{1.3}$~yr.
Thus, all the results presented in this paper pertain to
relatively massive and relatively young clusters (except
in \S4.3).

We find that Model~3, with mass-{\it in\/}dependent disruption,
provides a good match to the observed mass-age distribution
of the Antennae clusters.
The best-fitting exponents in this model are $\beta \approx
-2$ and $\gamma \approx -1$.
Models~1 and 2, with mass-dependent disruption fare much
worse.
Even with complete freedom to adjust several parameters,
these models never come close to matching the data.
While our detailed comparisons are based on models in 
which the formation rates of clusters are constant, our 
main conclusion, that the disruption rates do not
depend on mass, is valid even if the formation rates 
are variable.  
As a check on these results, we have also rederived the
luminosity function of the Antennae clusters.
After corrections for interstellar extinction, we find
a good fit to a pure power law, $\phi(L) \propto
L^{\alpha}$, with $\alpha \approx -2$.
The fact that the mass and luminosity functions are
nearly identical power laws ($\alpha \approx \beta$) is
a consequence of---and additional support for---the weak
or nonexistent correlations between masses and ages, i.e.,
the decomposition $g(M,\tau) \propto \psi(M)\chi(\tau)$.
We have also investigated recent claims that there are
bends or other features in the mass and luminosity
functions of the Antennae clusters (Fritze-von Alvensleben
1999; Mengel et al. 2005; Anders et al. 2007).
We show that these features are artifacts of the
%incomplete or inappropriate
ways the samples are defined
and do {\it not\/} reflect physical processes involved in
the formation and disruption of the clusters.

In an effort to understand the physical basis for our
simple---but doubtless approximate---model for $g(M,\tau)$,
we first note that the resulting mass function, $\psi(M)
\propto M^{-2}$, resembles that of molecular clouds in
nearby galaxies.
This indicates that the efficiency of star formation in
the clouds is roughly independent of their masses.
We also consider a variety of processes that could
disrupt the protoclusters and clusters:
removal of ISM by stellar feedback, continued stellar
mass loss, tidal disturbances by passing molecular clouds,
gravitational shocks during rapid passages near the
galactic bulge and through the galactic disk, orbital decay
into the galactic center caused by dynamical friction,
and stellar escape driven by internal two-body relaxation.
We suggest that the massive young clusters in the Antennae
galaxies are disrupted in the following approximate
sequence: (1) ISM removal, $\tau \la 10^7$~yr, (2) stellar
mass loss, $10^7 {\rm yr} \la \tau \la 10^8 {\rm yr}$, (3)
tidal disturbances, $\tau \ga 10^8$~yr.
We have argued on theoretical grounds that these processes
would operate at rates roughly independent of the masses
of the clusters, consistent with our empirically-based
decomposition $g(M, \tau) \propto \psi(M)\chi(\tau)$ (which,
however, is only well established for $\tau \la 10^8$~yr).
In combination, these processes plausibly account for the
observed decline of the age distribution, $\chi(\tau)
\propto \tau^{-1}$.
In the longer term, after 12 Gyr or so, the escape of stars
driven by two-body relaxation will preferentially disrupt
low-mass clusters and imprint a peak or turnover in the
evolved mass function at $M_p \approx (1-2) \times 10^5
M_{\odot}$, similar to that for old globular clusters in
the Milky Way and other galaxies.

How general is this picture?
In the Introduction, we mentioned some circumstantial evidence in
favor of its wide applicability:
the statistics of embedded clusters in the solar
neighborhood (Lada \& Lada 2003), the luminosities of
the brightest clusters in different galaxies (Larsen 2002;
Whitmore 2003; WCF07), the similarity of the mass spectra
of molecular clouds in different galaxies (Blitz et al.
2007), and the fact that the early disruption of
protoclusters and clusters ($\tau \la 10^8$~yr) is
driven mainly by internal processes that depend weakly, 
if at all, on the properties of their host galaxies 
(\S4 here).
Furthermore, from a detailed study of the mass-age
distributions of clusters in the LMC and SMC, we find
results nearly identical to those presented here
for the clusters in the Antennae (CFW09).
This is a strong test because the Antennae and the
Magellanic Clouds represent very different environments
for star and cluster formation: two large interacting
galaxies on one hand, two small, relatively quiescent
galaxies on the other hand.
Nevertheless, further tests of this picture in other
galaxies would be beneficial.
If this picture turns out to be generally valid, it
will mean that the main difference between populations
of young clusters is simply in the normalization of
$g(M, \tau)$ and hence in the overall formation rate.
From this perspective, the objects traditionally designated
open, populous, globular, or super clusters would belong
to a continuum, with mass and age as the most fundamental
variables, and would not require different formation
and/or disruption processes to account for their
observed properties.

The picture presented here has the virtues of simplicity
and possible universality.
It captures the salient properties of the observed
mass-age distribution, at least for the populations
of clusters we have studied so far.
Nevertheless, we do not expect our model $g(M, \tau)$
to be strictly universal.
It may break down outside its domain of validity in most
galaxies or inside this domain in some extreme environments,
such as those dominated by population III stars.
There may also be minor deviations from the model in some
normal galaxies, even within the domain of validity, caused
for example by variations in the formation rates and/or
differences in the disruption rates by passing molecular clouds.
However, these are only likely to affect $g(M, \tau)$ at 
intermediate ages ($10^8 {\rm yr} \la \tau \la 10^9{\rm yr}$), 
which will be difficult to detect observationally.
We see an analogy between, on one hand, the potentially universal 
stellar IMF, represented by a simple, approximate formula (first 
by Salpeter (1955) and then revised in subsequent studies based 
on more modern data) and, on the other hand, the potentially
universal mass-age distribution $g(M, \tau)$ for star clusters,
represented here by a simple, approximate formula. 
Both the stellar IMF and the cluster $g(M, \tau)$ are thought 
to be the outcome of several complex physical processes, in ways 
not yet fully understood, and neither is expected to hold exactly 
in all situations.
Even so, these functions provide compact and useful summaries 
of the observations and are thus natural focal points for
future studies of the formation and early evolution of
stars and clusters.

\acknowledgements
We thank Drs Bruce Elmegreen, Douglas Heggie,
Mark Krumholz, Christopher McKee,
Dean McLaughlin, and Francois Schweizer for helpful 
comments on this paper.
SMF acknowledges support from the Ambrose Monell Foundation and from
NASA grant AR-09539.1-A. and RC acknowledges support from
NASA grant GO-10402.11-A, awarded by the Space Telescope Science
Institute, which is operated by AURA, Inc., under NASA contract
NAS5-26555.  

\appendix

\section{COMPARISON WITH OTHER WORKS}

The mass and luminosity functions of star clusters in 
the Antennae galaxies derived in several other studies 
differ from the simple power laws presented here.
However, as we demonstrate in this appendix, the claimed 
peaks, bends, and other deviations from power laws all 
result from incompleteness in the samples of clusters 
rather than from any physical processes involved in their 
formation or disruption. 
Before we discuss any specific claims for features in the
mass and luminosity functions, however, it is worth 
considering the credibility of such claims in general.
Features that appear near the luminosity limit or
detection threshold should be considered with skepticism,
as they are almost always due to incompleteness.
Another common problem is selection criteria that 
fail to account for the fading of the clusters with age,
or that are linked in some way to measurement errors.
Finally, we note that it is inherently much more likely
that selection effects or measurement errors would convert
a simple underlying function like a power law into a more 
complicated one with extra features than the other way around.

These concerns are well founded in practice.  
Based on observations taken with the WF/PC1 on $HST$ 
(before the spherical aberration was corrected),
Fritze-von Alvensleben (1999) claimed that the mass 
function of the young clusters in the Antennae galaxies 
had a lognormal form, with a peak at $M \sim 10^5 M_{\odot}$,
like that of old globular clusters. 
However, the purported peak was near the detection
limit, and much deeper observations taken with the
WFPC2 (after the correction of spherical aberration), 
showed that it was simply an artifact of incompleteness, 
and that the mass function was in fact well represented 
by a power law, $\psi(M) \propto M^{-2}$, over the full
range of masses, $10^4 M_{\odot} \la M \la 10^6 M_{\odot}$ 
(ZF99 and \S3.2 here).

Mengel et al. (2005) have presented a mass function 
for the Antennae clusters that exhibits a gradual bend
at $M\sim10^5~M_{\odot}$.
Their result, derived from ground-based $K$-band
observations, is reproduced here as the open triangles in 
the top panel of Figure~15.
This is based on a sample that is limited at a constant 
luminosity with no restriction on age, instead of being 
limited at a variable luminosity, corresponding to a
constant mass, and restricted to relatively narrow 
intervals of age, as in our procedure.
Because the clusters fade with age, the Mengel et al.
sample does not include the same range of masses at all 
ages; older, low-mass clusters will be missed as they fade 
below the luminosity limit. 
We illustrate this in the bottom panel of Figure~15,
which shows a mass function constructed from our own sample
but with a selection procedure like that of Mengel et al.
Here, we have included clusters of all ages brighter than  
$K_{\rm lim} = 19$ (open circles),
where we have estimated the $K$-band magnitude of each
cluster from the observed $V$-band magnitude and the 
$V-K$ color predicted by the stellar population models 
of Bruzual \& Charlot (2003) for the age derived 
from our \textit{UBVI}H$\alpha$ observations. 
Evidently, this selection procedure introduces a bend
in the output mass functions similar to the one found 
by Mengel et al., even though the input mass function
is a power law (see \S 3.2 and Fig. 3).
We have also performed experiments with luminosities
limited in the $V$ and other bands and with gradual
and sharp cutoffs, and we always recover a similar,
artificial bend in the mass function.
Mengel et al. (2005) themselves appear to be aware of this
problem and urge caution in interpreting the bend in
their mass function as a real feature.

Anders et al. (2007) have derived a luminosity function 
of Antennae clusters with a peak near $M_V\approx-8.5$,
based on the same WFPC2 observations used here but a very 
different selection procedure.
In particular, they include only clusters with photometric 
errors $\leq0.2$~mag in \textit{UBVI}, thereby excluding many 
real clusters.
Since most of the missing clusters are faint, the luminosity
function derived in this way will fall below the true one at 
low luminosities, potentially introducing an artificial peak.
To test for this, we have repeated the steps in the Anders
et al. selection procedure, but using our own sample of 
clusters, for which the input luminosity function is known 
to be a power law (see \S 3.2 and Fig. 4).
In this case, the output luminosity function resembles the
one presented by Anders et al., with strong deviations from 
a power law, resulting primarily from the elimination of 
clusters with large uncertainties in the $U$-band photometry.
Thus, we conclude that the purported peak is an artifact 
of this selection bias.
A forthcoming study of the Antennae clusters based on 
deeper, higher-resolution observations taken with the 
ACS on {\em HST} confirms and extends the results 
presented in this paper, finding a luminosity function 
that is well represented by a pure power law with 
$\alpha\approx-2$ down to at least $M_V\approx-7$, 
with no evidence for a break or flattening near 
$M_V\approx-8.5$ (Whitmore et al. 2009).

\section{FORMULAE FOR $\bar{g}(M)$ and $\bar{g}(\tau)$}

Here, we present formulae for $\bar{g}(M)$ and $\bar{g}(\tau)$,
the averages of $g(M, \tau)$ over age ($\tau_1 \le \tau \le
\tau_2$) and mass ($M_1 \le M \le M_2$), respectively,
for the three disruption models discussed in Section~2.
The equations that follow were derived from equations~(6)
and (12)--(15).
The function $M_d(\tau)$ in equations~(B3) and~(B4) is the
inverse of the function $\tau_d(M)$ in equations~(B1) and~(B2).
The formulae for $\bar{g}(M)$ and $\bar{g}(\tau)$ are valid
for the stated values of the exponents $k$ and $\gamma$ in
the disruption models.
They are valid for all values of the exponent of the initial
mass function except $\beta = -1$, which is far enough from
the observed value $\beta \approx -2$ to be irrelevant in
the present context.

{\it Model~1: Sudden Mass-Dependent Disruption\/}.

\begin{equation}
\bar{g}(M) = C (M/M_*)^{\beta} \left\{
   \begin{array}{ll}
     0    &  \mbox{for $\tau_d(M) < \tau_1$}
\\
     R(M) &  \mbox{for $\tau_1 \le \tau_d(M) \le \tau_2$}
\\
     1    &  \mbox{for $\tau_2 < \tau_d(M)$}
   \end{array} \right.
\end{equation}

\begin{equation}
R(M) = \frac{\tau_d(M) - \tau_1}{\tau_2 - \tau_1}
\end{equation}

\begin{equation}
\bar{g}(\tau) = C_1 \left\{
   \begin{array}{ll}
     1       &  \mbox{for $M_d(\tau) < M_1$}
\\
     S(\tau) &  \mbox{for $M_1 \le M_d(\tau) \le M_2$}
\\
     0       &  \mbox{for $M_2 < M_d(\tau)$}
   \end{array} \right.
\end{equation}

\begin{equation}
S(\tau) = \frac{M_2^{\beta + 1} - M_d(\tau)^{\beta + 1}}
                {M_2^{\beta + 1} - M_1^{\beta + 1}}
\end{equation}

\begin{equation}
C_1 = \left(\frac{C}{\beta + 1}\right)
       \left(\frac{M_*}{M_2 - M_1}\right)
       \left[\left(\frac{M_2}{M_*}\right)^{\beta + 1}
       - \left(\frac{M_1}{M_*}\right)^{\beta + 1}\right]
\end{equation}

{\it Model~2: Gradual Mass-Dependent Disruption\/}.

\begin{equation}
\bar{g}(M) = C_2 (M/M_*)^{\beta + k}
              \left[T(M, \tau_2) - T(M, \tau_1)\right]
        \hspace{1cm} {\rm for} \,\,\,\,\, k \not= 0
\end{equation}

\begin{equation}
T(M,\tau) = \left[1 + k(M/M_*)^{-k}
             (\tau/\tau_*)\right]^{(\beta + 1)/k}
\end{equation}

\begin{equation}
C_2 = \left(\frac{C}{\beta + 1}\right)
       \left(\frac{\tau_*}{\tau_2 - \tau_1}\right)
\end{equation}

\begin{equation}
\bar{g}(M) = C_3 (M/M_*)^{\beta}
        \hspace{1cm} {\rm for} \,\,\,\,\, k = 0
\end{equation}

\begin{equation}
C_3 = \left(\frac{C}{\beta + 1}\right)
       \left(\frac{\tau_*}{\tau_2 - \tau_1}\right)
       \left\{\exp\left[(\beta+1)\left(\frac{\tau_2}{\tau_*}
       \right)\right] - \exp \left[(\beta+1)
       \left(\frac{\tau_1}{\tau_*}\right)\right]\right\}
\end{equation}

\begin{equation}
\bar{g}(\tau) = C_1 \exp\left[(\beta + 1)(\tau/\tau_*)\right]
        \hspace{1cm} {\rm for} \,\,\,\,\, k = 0
\end{equation}
\newpage

{\it Model~3: Gradual Mass-Independent Disruption\/}.

\begin{equation}
\bar{g}(M) = C_4 (M/M_*)^{\beta}
        \hspace{1cm} {\rm for} \,\,\,\,\, \gamma \not= -1
\end{equation}

\begin{equation}
C_4 = \left(\frac{C}{\gamma + 1}\right)
       \left(\frac{\tau_*}{\tau_2 - \tau_1}\right)
       \left[\left(\frac{\tau_2}{\tau_*}\right)^{\gamma + 1}
       - \left(\frac{\tau_1}{\tau_*}\right)^{\gamma + 1}\right]
\end{equation}

\begin{equation}
\bar{g}(M) = C_5 (M/M_*)^{\beta}
        \hspace{1cm} {\rm for} \,\,\,\,\, \gamma = -1
\end{equation}

\begin{equation}
C_5 = C \left(\frac{\tau_*}{\tau_2 - \tau_1}\right)
       \ln \left(\frac{\tau_2}{\tau_1}\right)
\end{equation}

\begin{equation}
\bar{g}(\tau) = C_1 (\tau/\tau_*)^{\gamma}
        \hspace{1cm} {\rm for} \,\, {\rm all} \,\, \gamma
\end{equation}

\clearpage

\begin{figure}
\plotone{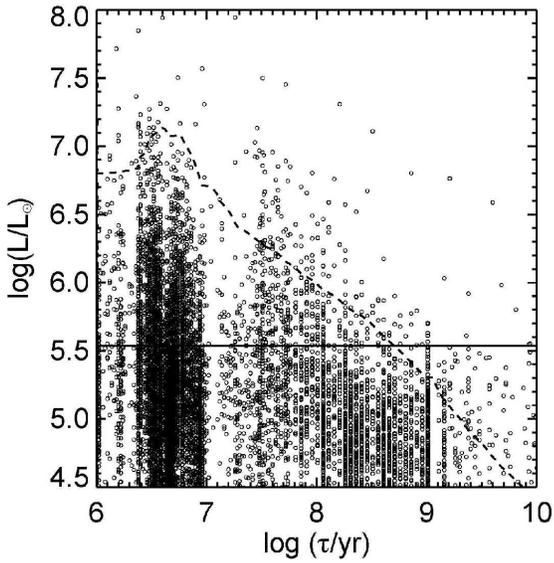}
\caption{Luminosity-age plane for the Antennae clusters.
The luminosities are in the $V$ band and have been
corrected for extinction.
The dashed diagonal line is the evolutionary track of
a model cluster with an initial mass of $M = 
2\times10^5$~$M_{\odot}$ (Bruzual \& Charlot 2003), 
while the solid horizontal
line at $L = 3\times10^5$~$L_{\odot}$ is the approximate
upper limit for stellar contamination.
The vertical gap in the data points just above $\tau = 1 
\times 10^7$~yr is an artifact caused by loops in the color
tracks of stellar population models during the RSG phase
(see the text).}
\end{figure}

\begin{figure}
\plotone{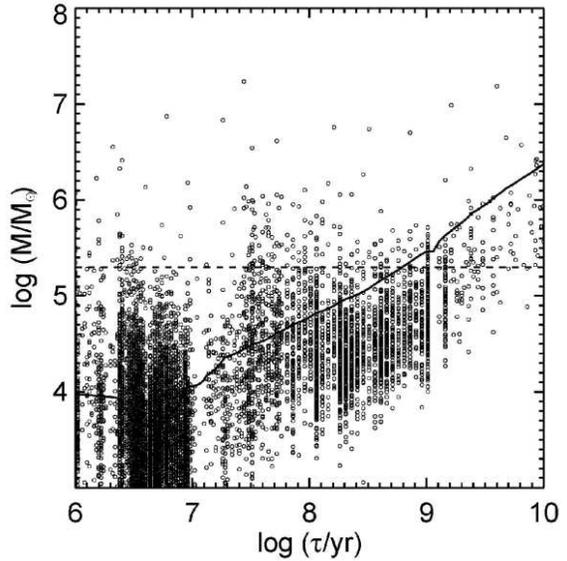}
\caption{Mass-age plane for the Antennae clusters.
This is an alternative, but equivalent, representation
of the data plotted in Fig.~1.
The lines in the two diagrams are also equivalent: the
dashed horizontal one indicates $M = 2\times 
10^5$~$M_{\odot}$, while the solid diagonal one indicates
$L = 3\times10^5$~$L_{\odot}$, the approximate upper limit
for stellar contamination.
The vertical gap in the data points just above $\tau = 1                                                              
\times 10^7$~yr is an artifact caused by loops in the color
tracks of stellar population models during the RSG phase
(see the text).}
\end{figure}

\begin{figure}
\plotone{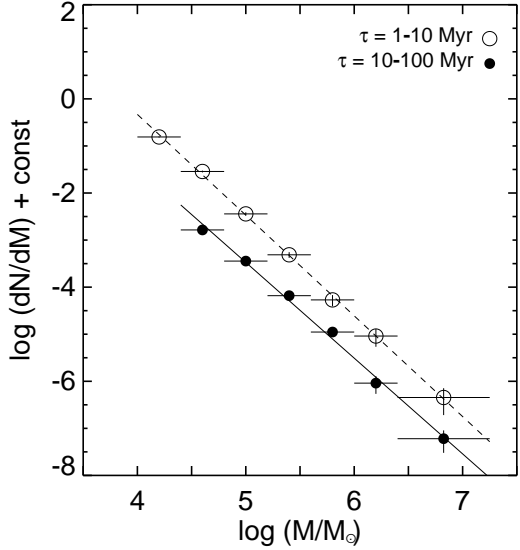}
\caption{Mass function of Antennae clusters in the
indicated intervals of age.
The lines are power laws, $dN/dM \propto M^{\beta}$,
with best-fit exponents $\beta=-2.14\pm0.03$ (dashed)
and $\beta=-2.03\pm0.07$ (solid) for the younger and
older clusters, respectively.}
\end{figure}

\begin{figure}
\plotone{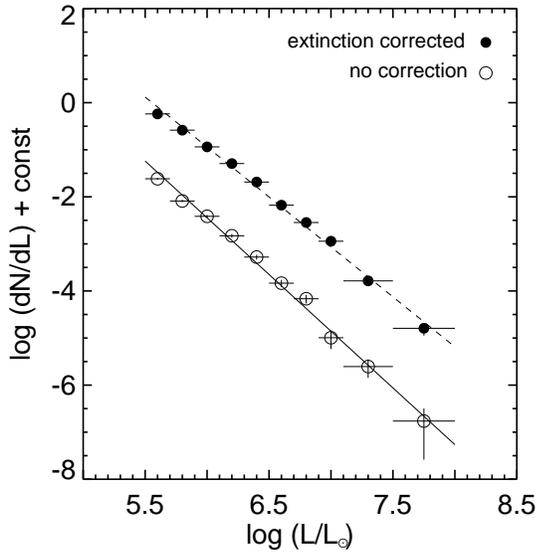}
\caption{Luminosity function of Antennae clusters in
the $V$ band, with and without corrections for extinction
(as indicated).
The lines are power laws, $dN/dL \propto L^{\alpha}$,
with best-fit exponents $\alpha=-2.09\pm0.05$ (dashed)
and $\alpha=-2.37\pm0.06$ (solid) for the corrected
and uncorrected luminosities, respectively.}
\end{figure}

\begin{figure}
\plotone{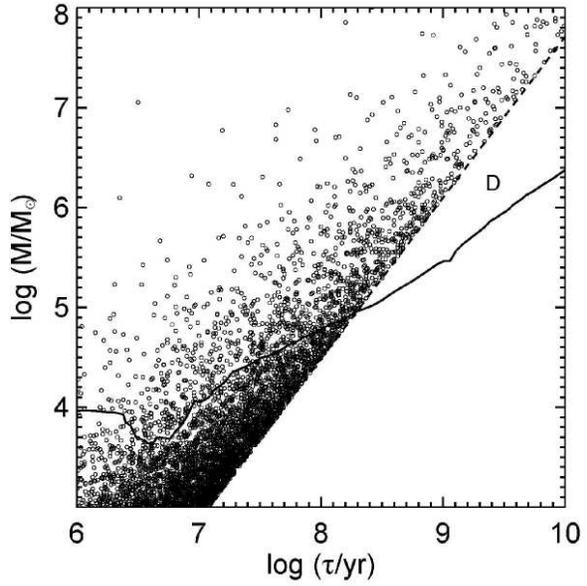}
\caption{Mass-age plane for a simulated population of clusters
based on Model~1.
This has a power-law initial mass function, a constant formation
rate, and sudden mass-dependent disruption with $\tau_*=5\times                                                       
10^7$~yr and $k=0.62$ in equation~(7).
The solid line is the approximate upper limit for stellar
contamination ($L = 3 \times 10^5 L_{\odot}$), while the
dashed line indicates the age at which clusters are destroyed.
The number of clusters in the simulated population above the
solid line is approximately the same as in the real Antennae
population.
The region marked D here is empty, in contrast to the same
region of Fig.~2.}
\end{figure}

\begin{figure}
\plotone{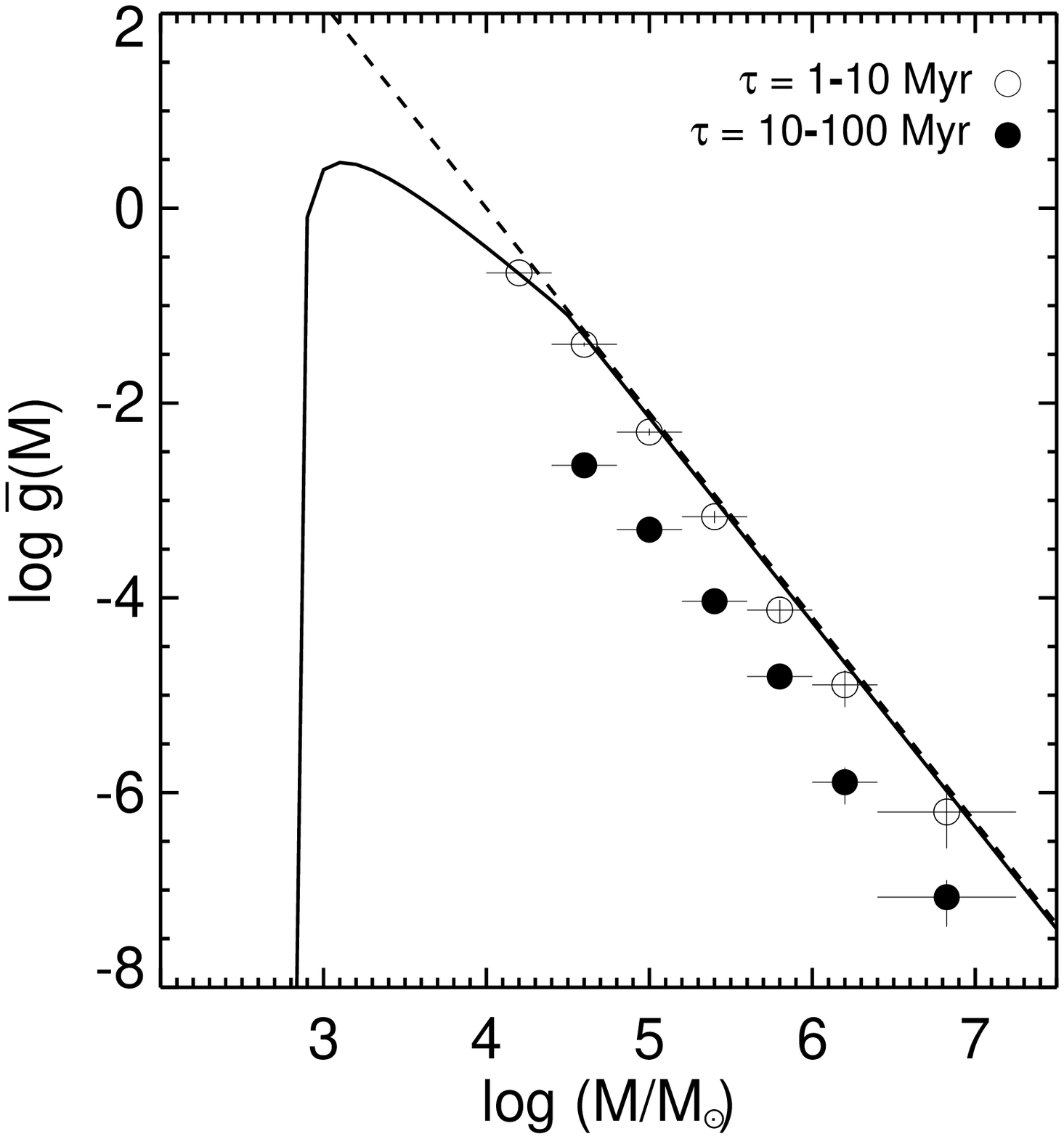}
\caption{Mass distribution averaged over the indicated intervals
of age for the Antennae clusters (data points) and for Model~1
(lines).
In this model, clusters have a power-law initial mass function,
a constant formation rate, and sudden mass-dependent disruption.
The dashed and solid lines were computed for the younger and
older clusters, respectively, from equations~(7), (B1), and (B2)
with $\tau_*=5\times10^7$~yr and $k=0.62$.}
\end{figure}

\begin{figure}
\plotone{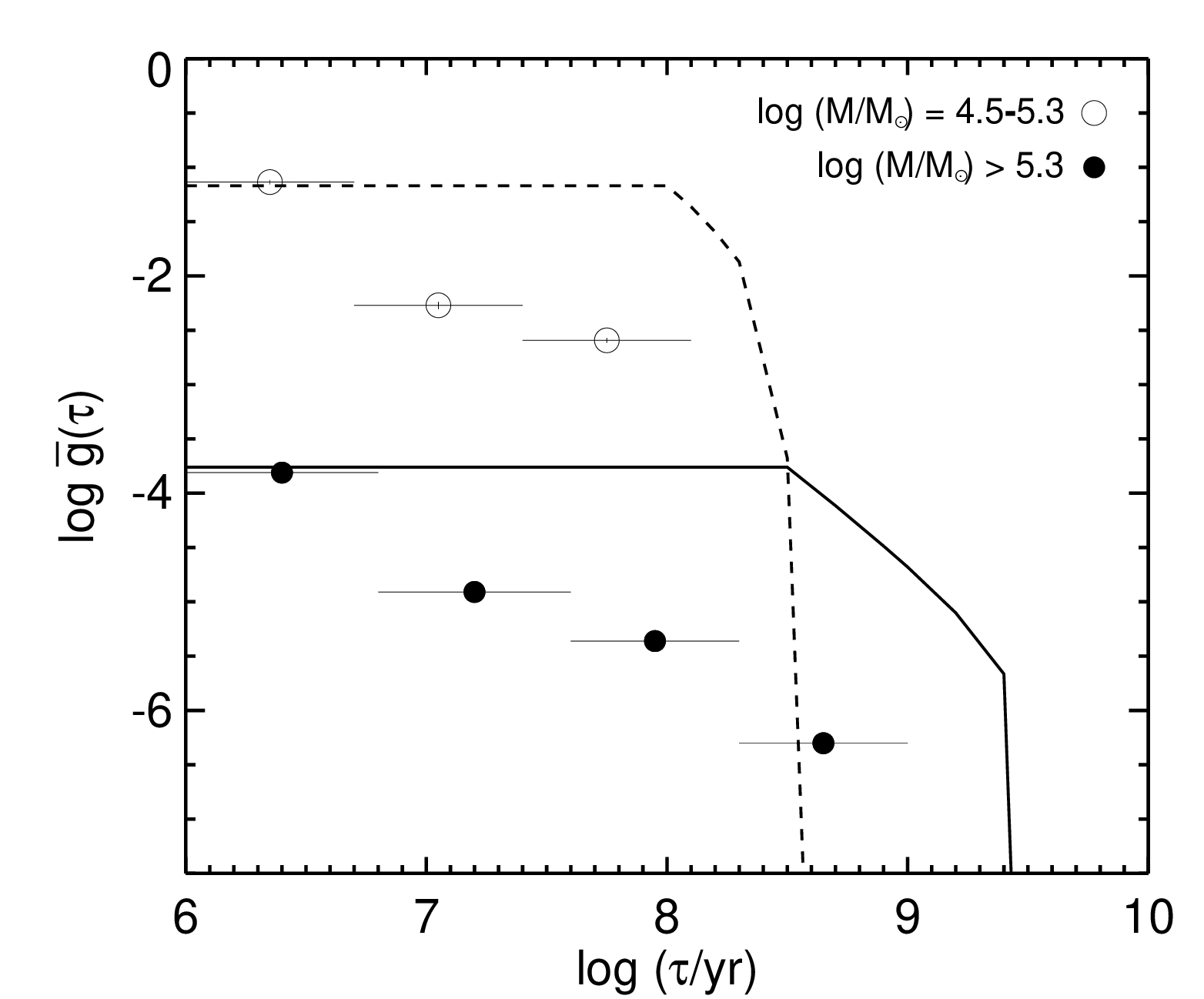}
\caption{Age distribution averaged over the indicated intervals
of mass for the Antennae clusters (data points) and for Model~1
(lines).
In this model, clusters have a power-law initial mass function,
a constant formation rate, and sudden mass-dependent disruption.
The dashed and solid lines were computed for less massive and
more massive clusters, respectively, from equations~(7) and
(B3)--(B5) with $\tau_*=5\times10^7$~yr and $k=0.62$.}
\end{figure}

\begin{figure}
\plotone{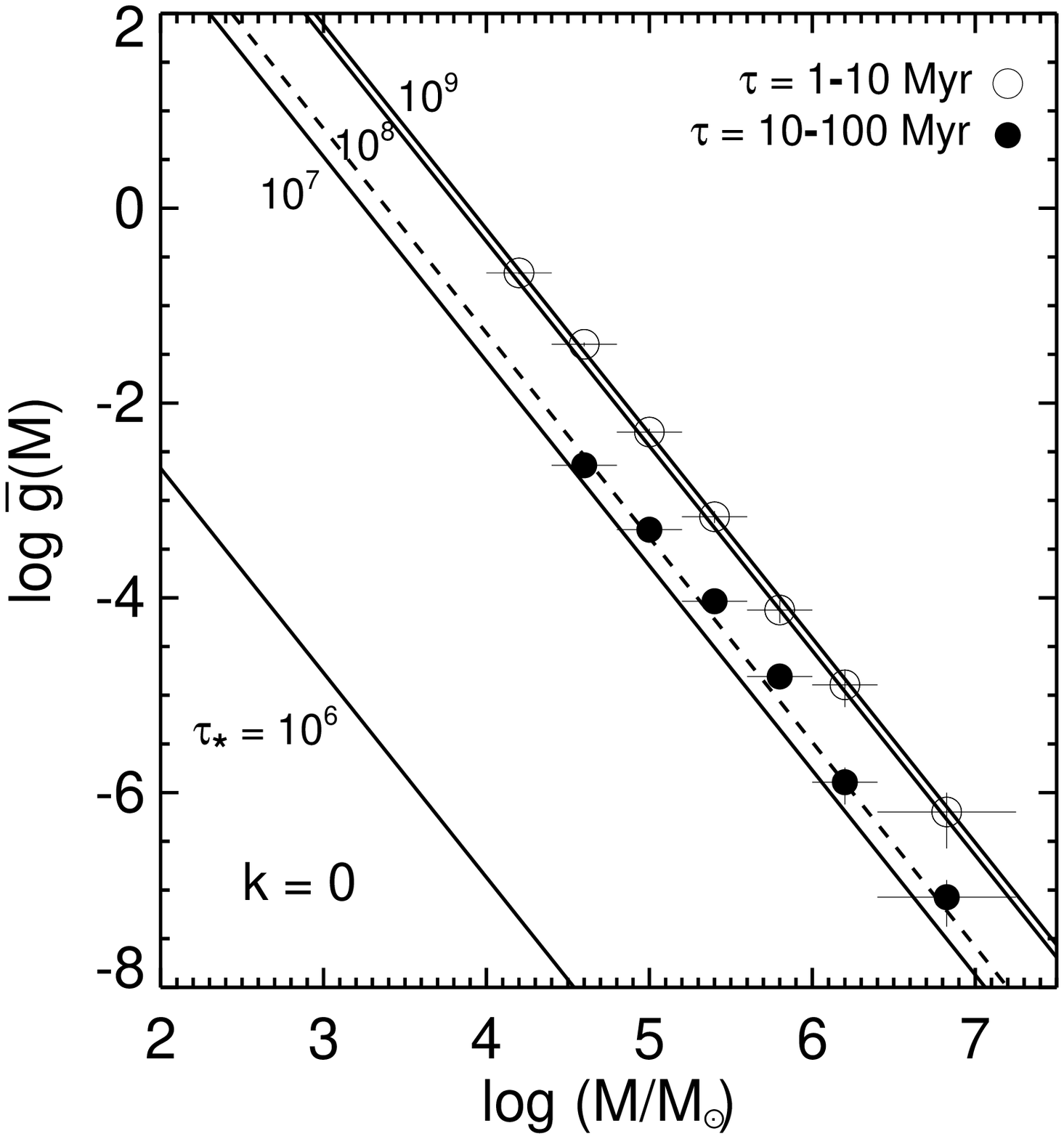}
\caption{Mass distribution averaged over the indicated intervals
of age for the Antennae clusters (data points) and for Model~2
with $k = 0$ (lines).
In this model, clusters have a power-law initial mass function,
a constant formation rate, and gradual mass-dependent disruption.
The solid lines were computed for the older clusters only from
equations~(B9) and (B10) with the indicated values of $\tau_*$.
The dashed line, with $\tau_* = 1.4 \times10^7$~yr, provides a
good fit to the observed distribution for the older clusters.}
\end{figure}

\begin{figure}
\plotone{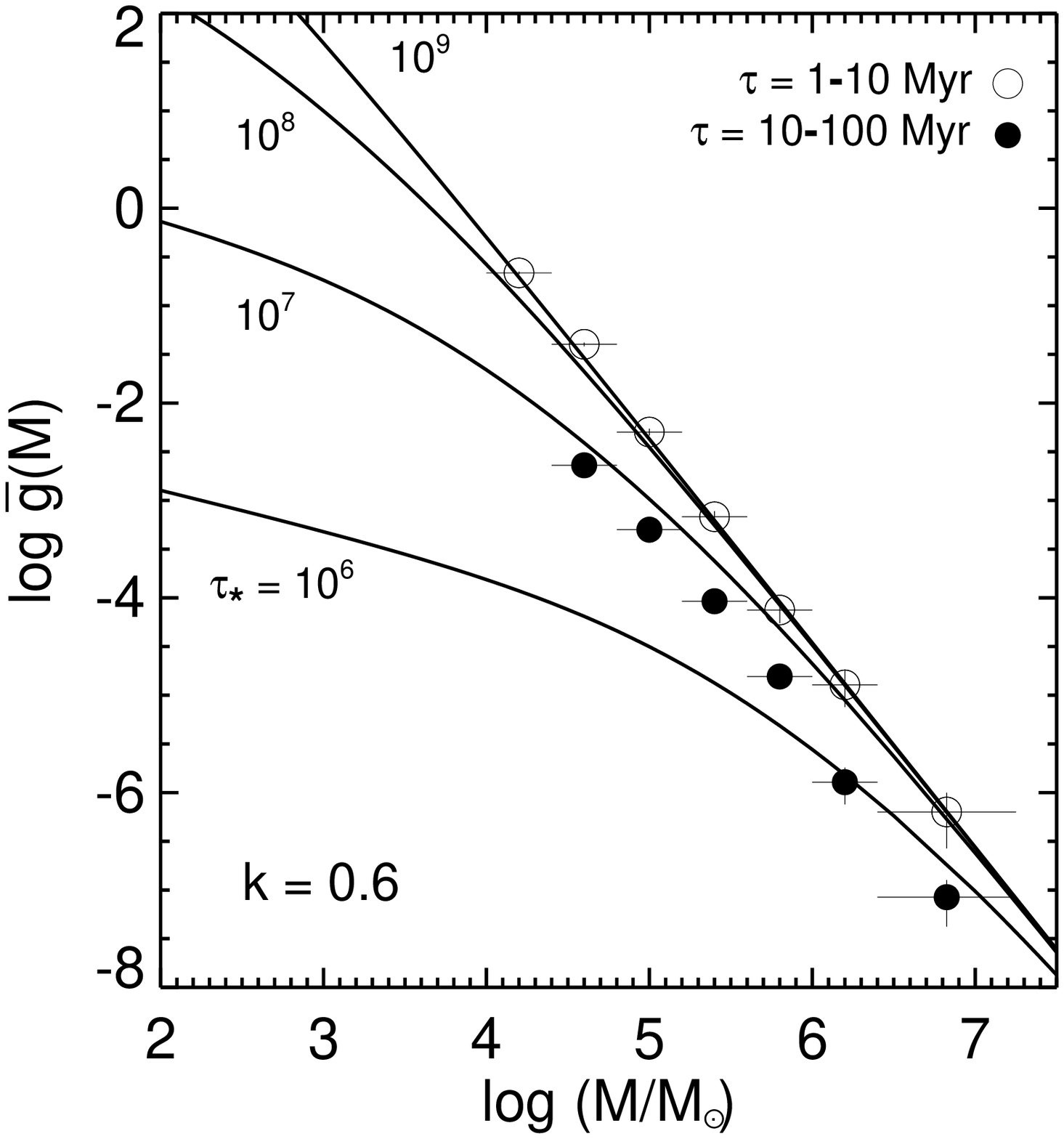}
\caption{Mass distribution averaged over the indicated intervals
of age for the Antennae clusters (data points) and for Model~2
with $k = 0.6$ (lines).
In this model, clusters have a power-law initial mass function,
a constant formation rate, and gradual mass-dependent disruption.
The solid lines were computed for the older clusters only from
equations~(B6)--(B8) with the indicated values of $\tau_*$.}
\end{figure}

\begin{figure}
\plotone{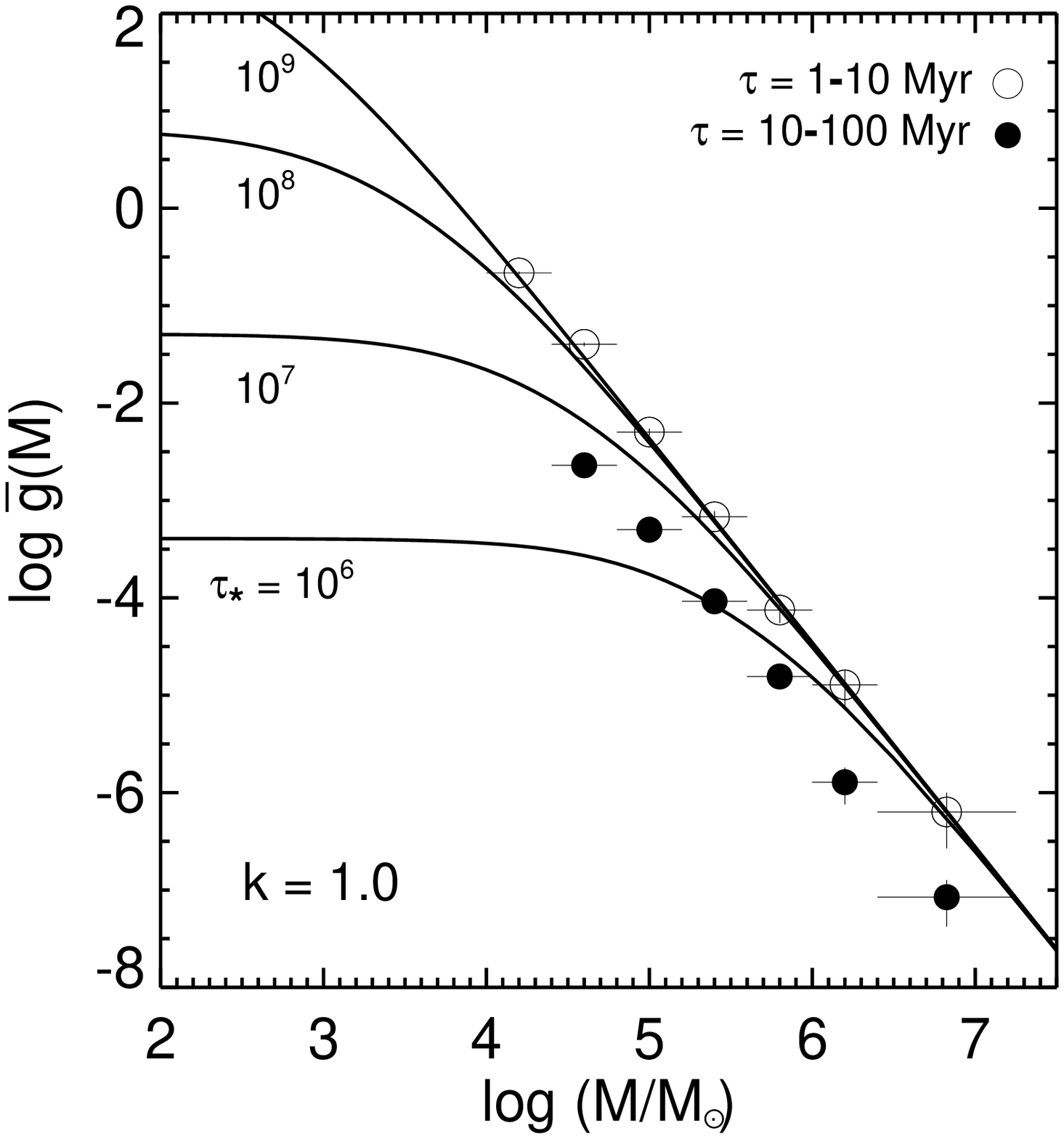}
\caption{Mass distribution averaged over the indicated intervals
of age for the Antennae clusters (data points) and for Model~2
with $k = 1.0$ (lines).
In this model, clusters have a power-law initial mass function,
a constant formation rate, and gradual mass-dependent disruption.
The solid lines were computed for the older clusters only from
equations (B6)--(B8) with the indicated values of $\tau_*$.}
\end{figure}

\begin{figure}
\plotone{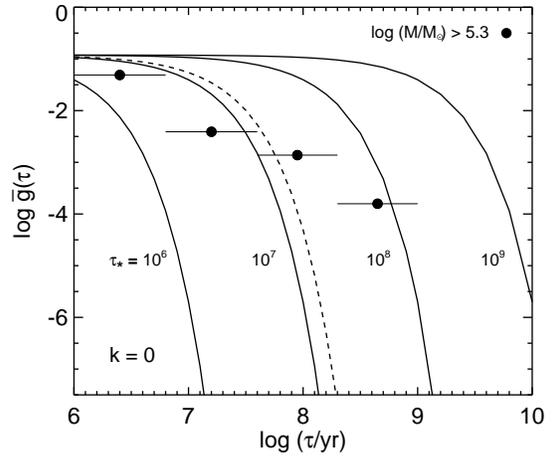}
\caption{Age distribution averaged over the indicated interval
of mass for the Antennae clusters (data points) and for Model~2
with $k = 0$ (lines).
In this model, clusters have a power-law initial mass function,
a constant formation rate, and gradual mass-dependent disruption.
The solid lines were computed from equations~(B5) and (B11) with
the indicated values of $\tau_*$.
The dashed line here corresponds to the dashed line in Fig.~8
(both with $\tau_* = 1.4 \times10^7$~yr).}
\end{figure}

\begin{figure}
\plotone{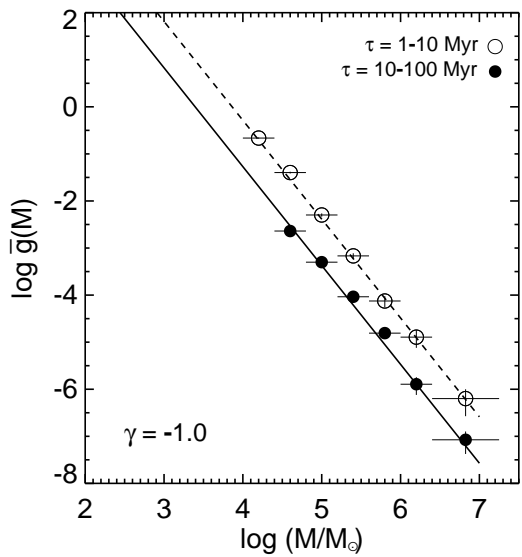}
\caption{Mass distribution averaged over the indicated intervals
of age for the Antennae clusters (data points) and for Model~3
with $\gamma = -1.0$ (lines).
In this model, clusters have a power-law initial mass function,
a constant formation rate, and gradual mass-{\it in\/}dependent
disruption.
The dashed and solid lines were computed for the younger and
older clusters, respectively, from equations (B14) and (B15).}
\end{figure}

\begin{figure}
\plotone{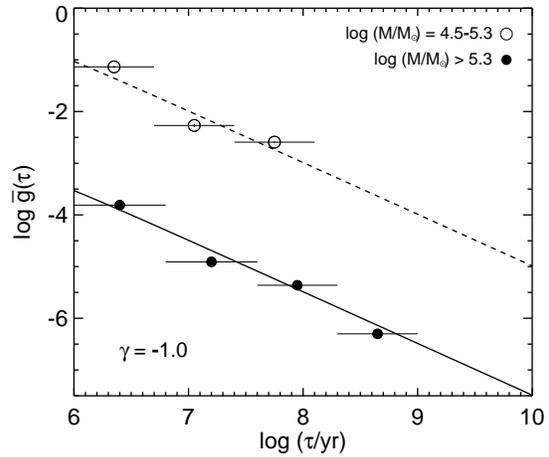}
\caption{Age distribution averaged over the indicated intervals
of mass for the Antennae clusters (data points) and for Model~3
with $\gamma = -1.0$ (lines).
In this model, clusters have a power-law initial mass function,
a constant formation rate, and gradual mass-{\it in\/}dependent
disruption.
The dashed and solid lines were computed for the less massive
and more massive clusters, respectively, from equations (B5)
and (B16).}
\end{figure}

\begin{figure}
\plotone{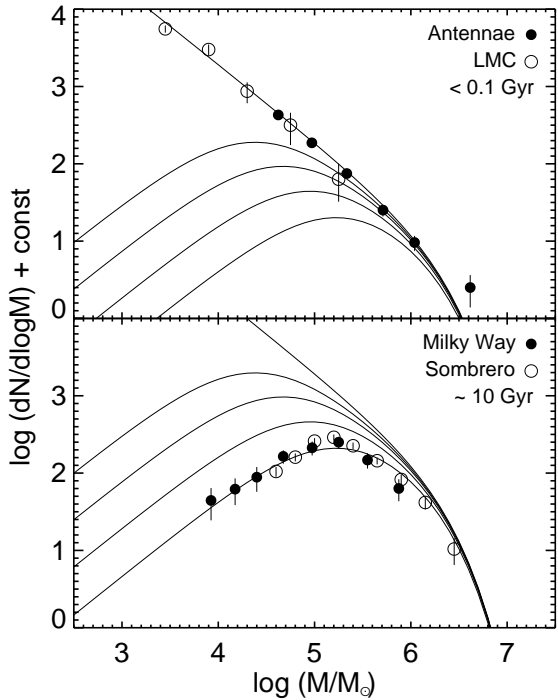}
\caption{
Mass functions of young clusters in the Antennae and LMC (upper panel) 
and  old globular clusters in the Milky Way and Sombrero (lower panel), 
as described in the text. The smooth curves in both panels are from 
the simple evaporation model, eqn. (18) with 
$\mu_{ev} = 2 \times 10^{-5} M_{\odot}{\rm yr}^{-1}$, at 
ages $\tau = 0$, 1.5, 3, 6, and 12 Gyr. 
Note the excellent fit of the model to the data at both $\tau = 0$ 
(upper panel) and $\tau = 12$~Gyr (lower panel).
}
\end{figure}

\begin{figure}
\epsscale{0.7}
\plotone{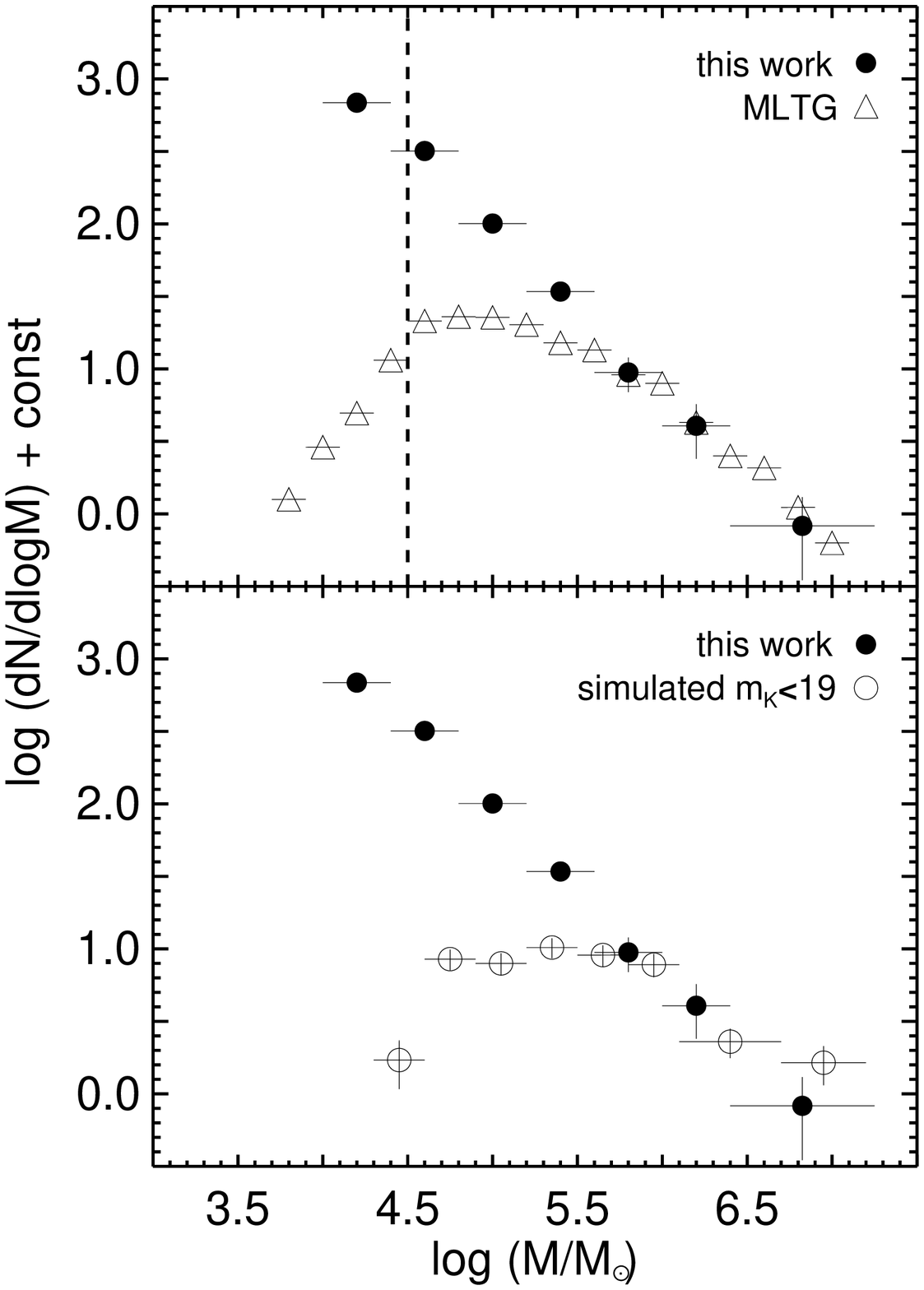}
\caption{{\it Upper Panel\/}: Mass function of Antennae clusters
from this work (filled circles) and from Mengel et al.
(2005) (open triangles).
The former is based on an age-restricted subsample ($\tau =$
1--10~Myr);
the latter is based on a subsample limited in $K_s$-band
luminosity but with no restriction on age.
The vertical dashed line is the stated 50\% completeness limit
for the Mengel et al. sample ($K_s \approx 18.4$).
{\it Lower Panel\/}: Mass function of Antennae clusters from this
work (filled circles) and for a subsample with a luminosity limit
similar to that of Mengel et al., based on $K_s$ magnitudes predicted from
stellar population models and the observed $V$ magnitudes
(open circles).}

\end{figure}


\begin{thebibliography}{}

\bibitem{ref1}
Anders, P., Bissantz, N., Boysen, L., de Grijs, R., \& Fritze-von Alvensleben, U. 2007,
MNRAS, 377, 91

\bibitem{ref1}
Baumgardt, H., Kroupa, P., \& Parmentier, G. 2008, MNRAS, 384, 1231

\bibitem{ref1}
Baumgardt, H., \& Makino, J. 2003, MNRAS, 340, 227

\bibitem{ref1}
Binney, J., \& Tremaine, S. 2008 Galactic Dynamics, 2nd Ed (Princeton: Princeton Univ. Press)

\bibitem{ref1}
Blitz, L., Fukui, Y., Kawamura, A., Leroy, A., Mizuno, N., \& Rosolowsky, E. 2007, 
in Protostars and Planets V, ed. B. Reipurth, D. Jewitt, \& K. Keil (Tucson: Univ. 
Arizona Press), 81

\bibitem{ref1}
Boutloukos, S. G., \& Lamers, H. J. G. L. M. 2003, MNRAS, 338, 717

\bibitem{ref1}
Bruzual, G., \& Charlot, S. 2003, MNRAS, 344, 1000

\bibitem{ref1}
Burkert, A., \& Smith, G. H. 2000, ApJ, 542, L95

\bibitem{ref1}
Calzetti, D., Kinney, A. L., \& Storchi-Bergmann, T. 1994, ApJ,
429, 572

\bibitem{ref1}
Chabrier, G. 2003, PASP, 115, 763

\bibitem{ref1}
Chandar, R., Fall, S. M., \& McLaughlin, D. E. 2007, ApJ, 668, L119

\bibitem{ref1}
Chandar, R., Fall, S. M., \& Whitmore, B. C. 2009, ApJ, submitted (CFW09)

\bibitem{ref1}
Chernoff, D. F., \& Weinberg, M. D. 1990, ApJ, 351, 121

\bibitem{ref1}
Dowell, J. D., Buckalew, B. A., \& Tan, J. C. 2008, AJ, 135, 823

\bibitem{ref1}
Elmegreen, B. G. 2002, ApJ, 564, 773

\bibitem{ref1}
Elmegreen, B. G., \& Efremov, Y. N. 1997, ApJ, 480, 235

\bibitem{ref1}
Elmegreen, B. G., \& Scalo, J. 2004, ARA\&A, 42, 211

\bibitem{ref1}
Fall, S. M. 2006, ApJ, 652, 1129

\bibitem{ref1}
Fall, S. M., Chandar, R., \& Whitmore, B. C. 2005, ApJ, 631, L133
(FCW05)

\bibitem{ref1}
Fall, S. M., \& Zhang, Q. 2001, ApJ, 561, 751

\bibitem{ref1}
Fitzpatrick, E. L. 1999, PASP, 111, 63

\bibitem{ref1}
Fleck, R. C. 1996, ApJ, 458, 739

\bibitem{ref1}
Fritze-von Alvensleben, U. 1999, A\&A, 342, L25

\bibitem{ref1}
Fukushige, T., \& Heggie, D. C. 1995, MNRAS, 276, 206

\bibitem{ref1}
Harris, W. E. 1996, AJ, 112, 1487

\bibitem{ref1}
Harris, W. E., \& Pudritz, R. E., 1994, ApJ, 429, 177

\bibitem{ref1}
Hills, J. G. 1980, ApJ, 225, 986

\bibitem{ref1}
Jord{\'a}n, A., et al. 2007, ApJS, 171, 101

\bibitem{ref1}
Kennicutt, R. C., Edgar, B. K., \& Hodge, P. W. 1989, ApJ, 337, 761

\bibitem{ref1}
Kroupa, P. 2001, MNRAS, 322, 231

\bibitem{ref1}
Kroupa, P., \& Boily, C. M. 2002, MNRAS, 336, 1188

\bibitem{ref1}
Krumholz, M. R., Matzner, C. D., \& McKee, C. F. 2006, ApJ, 653, 361

\bibitem{ref1}
Lada, C. J., \& Lada, E. A. 2003, ARA\&A, 41, 57

\bibitem{ref1}
Larsen, S. S. 2002, AJ, 124, 1393

\bibitem{ref1}
Leitherer, C. et al. 1999, ApJS, 123, 3

\bibitem{ref1}
Matzner, C. D., \& McKee, C. F. 2000, ApJ, 545, 364

\bibitem{ref1}
McKee, C. F., \& Ostriker, E. C. 2007, ARA\&A, 42, 211

\bibitem{ref1}
McKee, C. F., \& Williams, J. P. 1997, ApJ, 476, 144

\bibitem{ref1}
McLaughlin, D. E. 2000, ApJ, 539, 618

\bibitem{ref1}
McLaughlin, D. E., \& Fall, S. M. 2008, ApJ, 679, 1272

\bibitem{ref1}
Mengel, S., Lehnert, M. D., Thatte, N., \& Genzel, R. 2005, A\&A,
443, 41

\bibitem{ref1}
Mihos, J. C., Bothun, G. D., \& Richstone, D. O. 1993, ApJ, 418, 82 

\bibitem{ref1}
Parmentier, G., Goodwin, S. P., Kroupa, P., \& Baumgardt, H. 2008, ApJ, 678, 347

\bibitem{ref1}
Salpeter, E. 1955, ApJ, 121, 161

\bibitem{ref1}
Spitler, L. R., Larsen, S. S., Strader, J., Brodie, J. P., Forbes, D. A., \& Beasley, M. A. 
2006, AJ, 132, 1593

\bibitem{ref1}
Spitzer, L. 1958, ApJ, 127, 17

\bibitem{ref1}
Spitzer, L. 1987, Dynamical Evolution of Globular Clusters
(Princeton: Princeton Univ. Press)

\bibitem{ref1}
Takahashi, K., \& Portegies Zwart, S. F. 2000, ApJ, 535, 759

\bibitem{ref1}
V{\'a}zquez-Semadeni, E., Ballesteros-Paredes, J., \& Rodr{\'i}guez, L. F. 1997, ApJ, 474, 292

\bibitem{ref1}
Vesperini, E., \& Zepf, S. E. 2003, ApJ, 587, L97

\bibitem{ref1}
Vesperini, E., Zepf, S. E., Kundu, A., \& Ashman, K. M. 2003, ApJ, 593, 760

\bibitem{ref1}
Wada, K., Spaans, M., \& Kim, S. 2000, ApJ, 540, 797

\bibitem{ref1}
Whitmore, B. C. 2003, in A Decade of Hubble Space Telescope Science,
ed. M. Livio, K. Noll, \& M. Stiavelli (Cambridge: Cambridge Univ. Press), 153

\bibitem{ref1}
Whitmore, B. C. 2004, in ASP Conf. Ser. 322, The Formation and Evolution
of Massive Young Star Clusters, ed. H. J. G. L. M. Lamers, L. J. Smith, \& A. Nota
(San Francisco: ASP), 411

\bibitem{ref1}
Whitmore, B. C., Chandar, R., \& Fall, S. M. 2007, AJ, 133, 1067 (WCF07)

\bibitem{ref1}
Whitmore, B. C., \& Zhang, Q. 2002, AJ, 124, 1418

\bibitem{ref1}
Whitmore, B. C., Zhang, Q., Leitherer, C., Fall, S. M., Schweizer, F.,
\& Miller, B. W. 1999, AJ, 118, 1551

\bibitem{ref1}
Whitmore, B. C., et al. 2009, AJ, submitted

\bibitem{ref1}
Wilson, C. D., Scoville, N., Madden, S. C., \& Charmandaris, V. 2003, ApJ, 599, 1049

\bibitem{ref1}
Zhang, Q., \& Fall, S. M. 1999, ApJ, 527, L81 (ZF99)


\end{thebibliography}
\end{document}